\documentclass[%
 reprint,
 superscriptaddress,
 amsmath,amssymb,
 aps, floatfix
]{revtex4-2}
\setcitestyle{super}
\usepackage{dcolumn}% Align table columns on decimal point
\newcolumntype{.}{d{.}{.}{-1}}

\usepackage{bm}% bold math
\usepackage{booktabs}

\usepackage[latin1]{inputenc}

\usepackage{mathtools}
\usepackage{ amssymb }
\let\vec\mathbf

\usepackage[shortlabels]{enumitem}

\usepackage{physics}
\usepackage{hyperref}
\usepackage[htt]{hyphenat}

\usepackage{svg} % use svg
\usepackage{float} % fix position of figures
\usepackage{graphicx}

\usepackage{placeins} % for float barrier

\graphicspath{ {./figures/} }

\begin{document}
%TODO
% what are the correct affiliations

\title{From quantum alchemy to Hammett's equation:
Covalent bonding from  atomic energy partitioning}
% \title{Simple covalent bonding trends via atomic energy decomposition based on quantum alchemy}
% \title{From quantum alchemy to Hammett's equation:
% Revisiting the covalent bond}
%\title{Different view on the chemical bond in chemical space}
%\title{Quantum alchemy based chemical bonding}
%\title{Fresh views of chemical bonding from computational alchemy}
%\title{Understanding covalent bonding with computational alchemy}
% How covalent bonds transform in chemical space
% The covalent bond through chemical space
% A quantum alchemy view on the covalent bond 
% Bonding trends in chemical space
% Relationships between covalent bonds of different elements

% no passive tone
% no alchemy
% not to general

\author{Michael J. Sahre}
\affiliation{University of Vienna, Faculty of Physics, Kolingasse 14-16, 1090 Vienna, Austria}
\affiliation{University of Vienna, Vienna Doctoral School in Chemistry (DoSChem), W\"ahringer Str. 42, 1090 Vienna, Austria.}
\author{Guido Falk von Rudorff}
\affiliation{University Kassel,
Department of Chemistry, 
Heinrich-Plett-Str.40,34132 Kassel, Germany}
%\affiliation{University of Vienna, Faculty of Physics, Kolingasse 14-16, 1090 Vienna, Austria}
% \affiliation{Institute for Pure and Applied Mathematics (IPAM), University of California, Los Angeles, 460 Portola Plaza, Los Angeles, CA 90095.}
%\affiliation{Institute of Physical Chemistry and National Center for Computational Design and Discovery of Novel Materials (MARVEL), Department of Chemistry, 
%University of Basel, 
%Klingelbergstr. 80,
%4056 Basel, Switzerland}
\author{O. Anatole von Lilienfeld}
\email{anatole.vonlilienfeld@utoronto.ca}
\affiliation{Vector Institute for Artificial Intelligence, Toronto, ON, M5S 1M1, Canada}
\affiliation{Departments of Chemistry, Materials Science and Engineering, and Physics, University of Toronto, St. George Campus, Toronto, ON, Canada}
\affiliation{Machine Learning Group, Technische Universit\"at Berlin and Institute for the Foundations of Learning and Data, 10587 Berlin, Germany}

\begin{abstract}
%The concept of chemical bonds plays a fundamental role for our atomistic understanding of the behavior of matter. 
%Intuitive models of chemical bonding have played an important role in the development of the chemical sciences. 
We present an intuitive and general analytical approximation estimating the energy of covalent single and double bonds between participating atoms in terms of their respective nuclear charges with just three parameters,
$[{E_\text{AB} \approx a - b Z_\text{A} Z_\text{B} + c (Z_\text{A}^{7/3} + Z_\text{B}^{7/3})}]$. 
The functional form of our expression models an alchemical atomic energy decomposition between participating atoms A and B. 
After calibration, reasonably accurate bond energy estimates are obtained for hydrogen-saturated diatomics composed of
$p$-block elements coming from the same row $2\le n\le 4$ in the periodic table. 
Corresponding changes in bond energies due to substitution of atom B by C can be obtained via simple formulas. 
While being of different functional form and origin, 
our model is as simple and accurate as Pauling's well-known electronegativity model. 
Analysis indicates that the model's response in covalent bonding to variation in nuclear charge is near-linear---which is 
consistent with Hammett's equation.
\end{abstract}

% line 100 (Corresponding changes...) 

% "The model
% accurately reproduces single bond energies with an MAE of 1.8 kcal/mol for saturated diatomic
% molecules composed of any chemical main group elements coming from the same period n = 2, 3, or 4."
\maketitle

\section{Introduction}
%The classics, why do we start with: Due to their...
Due to their direct link to thermodynamics stability, computational predictions of binding trends among molecules and materials have greatly improved %experimental
design choices in the chemical sciences.\cite{lilie_design, Honarparvar2014, Kumalo2015, London2014, Norskov2009, Curtarolo2013, Hannagan2021, Schwalbe-Koda2021}
%In particular, stability predictions are important to identify experimentally accessible compounds.
%A dominating factor with respect to stability is chemical bonding. 
Unfortunately, while numerical bonding estimates obtained from modern and computationally demanding quantum methods are very accurate and predictive, they are at the same time difficult to grasp with human intuition due to the inherent complexity of their solutions to the electronic Schr\"odinger equation.\cite{purple_book}
% More intuitive models popularized in undergraduate text-books, such as H\"uckel-theory\cite{huckel} or Pauling's electronegativity\cite{pauling}, are based on more approximate formulations of quantum mechanics 
% in terms of Molecular Orbital\cite{Murrell_MO, mulliken_MO} theory and Valence Bond CITE[HeitlerLondon,\cite{pauling}], respectively. 
Already very approximate models, however, can yield intuitive  descriptions of important bonding features. 
For example, Lewis' simple concept of binding electron pairs, in conjunction with the Aufbau principle, imposes relevant constraints from Pauli's exclusion principle,\cite{FrenkingOnLewis}
% For example, Lewis' simple concept of binding electron pairs is consistent with 
% Pauli's exclusion principle combined with writing the wavefunction as a sum of localized orbital products.
% this deceptive according to this paper from Frenking:
%  A critical look at linus pauling's influence on the understanding of chemical bonding
or the delocalization of $\pi$-electrons can be easily understood within H\"uckel theory (FIG 1,left).\cite{huckel}
Pauling's model for covalent bond energies is based on the idea to decompose the wavefunction within valence bond theory into ionic and covalent terms, leading to the chemically intuitive concept of electronegativity (FIG 1, center).\cite{pauling} 
% (there is new work from Roald Hoffmann on electronegativity but based in Allens EN not Paulings; they state that electronegativity can be used to guide chemical design so it is related to our motivation, also their EN is defined via reaction energies so it is related to chemical bond energies)
% "A model proposed by Pauling" instead of "Pauling's model..." because this equation is not known as Paulings model
Because of their universality due to their foundation in quantum mechanics, these basic models are applicable across chemical compound space, and have proven powerful for advancing our understanding of chemistry--despite their approximate nature.
% Less approximate and more recently, energy decomposition methods have been introduced to provide a more detailed understanding of chemical bonding.
% Some methods decompose the energy into different physical contributions, thers partition it onto atoms.
% Partitioning of quantum mechanical observables onto the constituting parts of the quantum many-body system can be done in arbitrarily many ways. Also within the framework of computational alchemical perturbation density functional theory~\cite{VonRudorff2020} one can meaningfully quantify the effect of different binding partners on the atomic energy of every atom in the system~\cite{VonRudorff2019}.

Less approximate and more recently, energy decomposition methods have been introduced to provide a more detailed understanding of chemical bonding. Partitioning of quantum mechanical observables onto the constituting parts of the quantum many-body system can be done in arbitrarily many ways.
Some methods decompose the energy into different physical contributions~\cite{Morokuma, Ziegler1977,frenking_bickelhaupt,Rahm2015, Keith_decomposition}, others partition it onto atoms~\cite{IQA, Martin_atomic_energy, Bonds1995, Eriksen2020, VonRudorff2020}. These methods have for example been useful to explain differences in interatomic interactions,~\cite{Frenking2022} stability~\cite{Blokker2021, Hansen2022} or torsional energy profiles~\cite{Darley2008} across various molecules.

Also within the framework of computational alchemical perturbation density functional theory~\cite{VonRudorff2020} one can meaningfully quantify the effect of different binding partners on the atomic energy of every atom in the system~\cite{VonRudorff2019}. 
%and is as such a suitable starting point for an investigation of binding energy trends as a function of composition.
Here, we introduce a model of covalent bonding which we compare to molecular orbital (MO) and valence bond (VB) theory in FIG 1. The model directly emerges from quantum alchemy (QA) based atomic energy decomposition arguments. It is, to the best of our knowledge new, yet as simple and accurate as Pauling's electronegativity model and of a distinctly different functional form and origin. We thus believe that it enables a fresh and intuitive perspective on covalent bonding.

\begin{figure*}[!ht]
\centering
\includegraphics[width=1.0\textwidth]{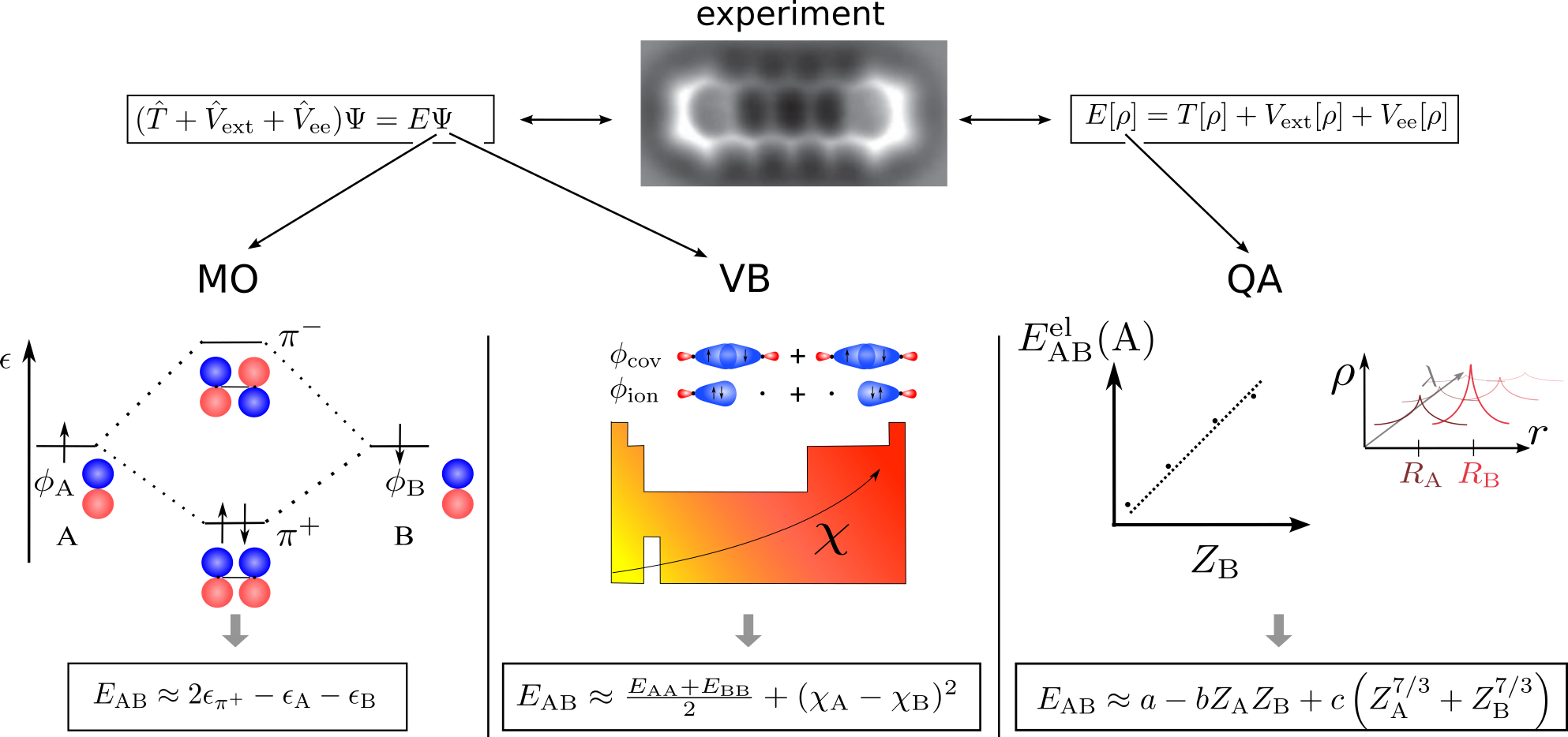}
\caption{
Experimental measurements (illustrated by atomic force microscopy image of pentacene molecule adapted from \citenum{AFM}. Reprinted with permission from AAAS.) inform 
exact theory (Schr\"odinger equation \& density functional theory) which informs
three intuitive approximate views of the chemical bond (MO, VB, QA) in chemical compound space.
MO represents molecular orbital theory accounting for bonding in energy diagrams (left).
VB corresponds to valence bond theory enabling the decomposition of the wavefunction into covalent and ionic parts (mid).
Quantum Alchemy (QA) enables direct partitioning into atomic energies based on thermodynamic integration over varying nuclear charges. 
MO, VB, and QA can be used to account for bonding trends.%\cite{Erwin,Walter}
}
\label{fig:conceptual}
\end{figure*}

Within quantum alchemy, two systems can be interconverted by either interpolating the Hamiltonians or treating a change in the system perturbatively.\cite{huckel2,Wilson1962, VonLilienfeld2013,VonRudorff2020} We connect the energies $E^\text{tot}$ and $E^\text{UEG}$ of a molecule and the uniform electron gas  via alchemical thermodynamic integration.\cite{VonRudorff2019}
Using the Hellmann-Feynman theorem\cite{HF1939} and the chain-rule affords a formally exact atomic energy $E_I$ partitioning with respect to nuclear charges, 
\cite{VonRudorff2019}
\begin{equation}
\begin{split}
    &E^\text{tot} - E^\text{UEG}\\ &= \sum_I \underbrace{Z_I\left(  \int d\vec{r} \frac{\int_0^1 d\lambda \rho(\lambda, \vec{r})}{|\vec{r}-\vec{R}_I|}+\frac{1}{2}\sum_{J \neq I} \frac{Z_J}{|\vec{R}_J-\vec{R}_I|}\right)}_{:= E_I},
\end{split}
\label{eq:atomic_energy}
\end{equation}
with $\rho(\lambda, \vec{r})$, $Z_I$, and $\vec{R}_I$ corresponding to 
the electron density, the nuclear charge, and the position of the nucleus, respectively. 

% Our DFT based calculated binding energies for varying nuclear charges connecting dozens of chemical bonds among $p$-block elements reveal simple dependencies of atomic binding contributions on nuclear charges.
Application of our alchemical decomposition scheme to binding energies between $p$-block elements reveals simple dependencies of atomic binding contributions on nuclear charges.
This has motivated us to introduce the following approximate Ansatz for the binding energy $E_\text{AB}$ between atoms A and B,
\begin{eqnarray}
    E_\text{AB} & \approx & \underbrace{\vphantom{a}a}_{{\text{period's offset}}} -  b\underbrace{\vphantom{\left(Z_\text{A}^{7/3}\right)}Z_\text{A} Z_\text{B}}_{\quad\qquad\mathllap{\propto\,\text{nuc. rep.}}} + \underbrace{c\left(Z_\text{A}^{7/3} + Z_\text{B}^{7/3}\right)}_{\mkern-140mu\mathrlap{\propto\, \text{free atoms}}}, \nonumber\\
    \label{eq:SRL*}
\end{eqnarray}
%nuclear charges $Z_\text{A}, Z_\text{B}$ of the mutually bonded atoms, 
with simple interpretation for each term and requiring just three global parameters $a, b, c$, which effectively account for interatomic distance and bond order (see Methods A for details).

%For details on the reasoning, the interested reader is referred to the Methodology section.
After calibration of parameters, we find that this simple model reproduces covalent binding among $p$-block elements of either the second, third, or fourth row of the periodic table reasonably well. However, we note that the model is only applied to systems A-B with remaining valencies saturated with hydrogens, i.e. further environmental influences are not yet being studied in this paper.
% We study the performance of this simple model for compounds A-B with remaining valencies saturated with hydrogens.
% After calibration of parameters, covalent binding among $p$-block elements of either the second, third, or fourth row of the periodic table is reproduced reasonably well.
In the following we also analyse and compare the model to density functional theory, quantum machine learning, semi-empirical and post-Hartree-Fock quantum mechanics,  Pauling's electronegativity model, and Hammett's equation.
% We subsequently discuss limitations, and formulate a generalized expression that predicts bonding trends across multiple periods.
We subsequently discuss limitations, extensions to double bonds and different electronic configurations and formulate a generalized expression that predicts bonding trends across multiple periods.

% \section{Numerical results}
\section{Results and Discussion}
\subsection{Performance}
\begin{figure*}[ht]
\centering
\includegraphics[width=1.0\textwidth]{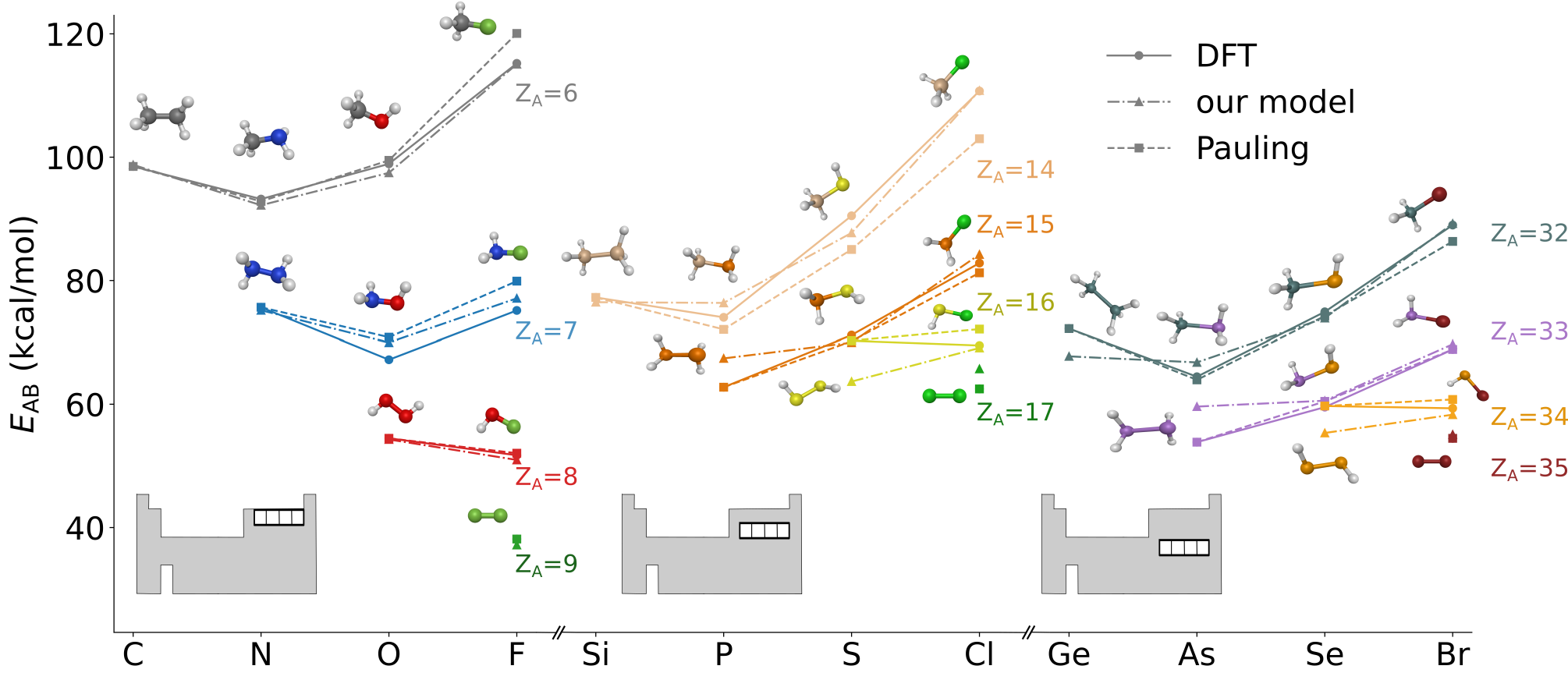}
\caption{Calculated bond dissociation energies from density functional theory (DFT, PBE0/def2-TZVP), our quantum alchemy based chemical bond model (Eq.~\eqref{eq:SRL*}), and Pauling's electronegativity model 
in Eq.~\eqref{eq:pauling}.
MAEs with respect to DFT amount to 1.8 and 1.4 kcal/mol for our and Pauling's model, respectively.}
\label{fig:bde}
\end{figure*}

After regression of parameters to DFT reference data, our model (Eq.~\eqref{eq:SRL*}) makes 
surprisingly accurate estimates.
In particular,
FIG.~\ref{fig:bde} shows calculated bond dissociation energies (BDEs) for homolytic cleavage of diatomics A-B saturated with Hydrogens,
\begin{equation}
    \text{H}_x \text{A} - \text{B} \text{H}_y \rightarrow \text{H}_x \text{A}^{\bullet} + ^{\bullet}\text{B} \text{H}_y. 
\end{equation}
Here, atoms A and B are fourth to seventh main group elements, both of either second, third or fourth row of the periodic table.
BDEs for calibration were obtained from density functional theory (see Computational Details) and the row dependent model parameters were determined from a least-square fit to the ten diatomics in each row (see TABLE~\ref{tab:coeff}).
The residual deviation of estimated binding energies  from the DFT reference amounts to an overall mean absolute error (MAE) of just 1.8\,kcal/mol across all rows, falling just short of the highly coveted `chemical accuracy' threshold of $\sim$1 kcal/mol. 
While such accuracy is extremely promising for such a simple functional form and so few parameters, we note the severe restrictions and limitations including fixed effective equilibrium geometries, participating elements, or bond orders.

\begin{table}
\centering
\caption{Coefficients and MAEs of our model Eq. \eqref{eq:SRL*} for  different rows $n$. $b$ is scaled such that $b Z_\text{A} Z_\text{B}$ is given in kcal/mol if $Z_\text{A}$ and $Z_\text{B}$ are given in atomic units.}
\begin{tabular}{ c c c c c}
 $n$ & $a$ (kcal/mol) & $b$ (630/$a_0$) & $c$ (kcal/mol) & MAE (kcal/mol) \\
 \hline
2 & \ \ 215.7 & 10.470 & 1.987 & 1.0 \\
3 & \ \ 392.7 & \ \ 8.823 & \ 1.496 & 2.4 \\
4 & 1109.2 & \ \ 6.180 & \ 0.813 & 2.2 \\
\end{tabular}
\label{tab:coeff}
\end{table}

\begin{figure}[ht]
\centering
\includegraphics[width=0.5\textwidth]{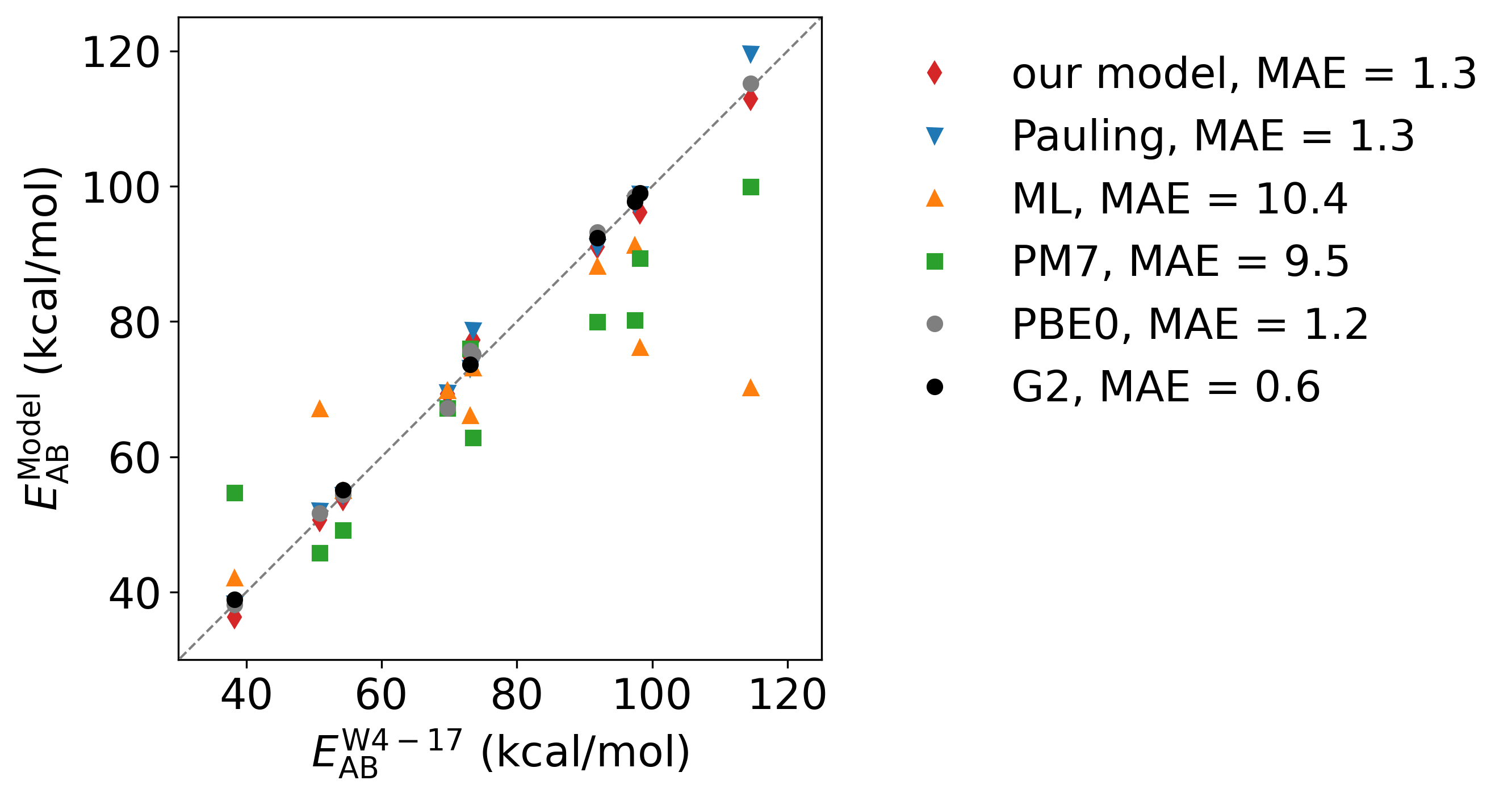}
\caption{Calculated BDEs using our model, Pauling's, various quantum methods (PM7, DFT (PBE0/DEF2-TZVP), G2) and a quantum machine learning model. Binding energies are scattered against the W4 entries\cite{W4} for ten saturated single bond diatomics composed of carbon, nitrogen, oxygen or fluorine.}
\label{fig:compare_models}
\end{figure}

To set this performance  into a wider perspective we compare calculated BDEs among other models well established in the literature.  
For this purpose, we selected all ten single bond diatomics from the second row contained within the legacy quantum chemistry W4-17 dataset\cite{W4} which provides highly accurate bond dissociation energies using explicit electron correlation methods for saturated diatomics in the second row.
FIG.~\ref{fig:compare_models} 
shows a scatter plot of BDEs obtained from various models.  
Calibrating our model using the W4-17 data yields a leave-one-out prediction error of 1.3 kcal/mol (see Computational Details).
Training a chemically agnostic quantum machine learning\cite{Huang2021,Tkatchenko2020,VonLilienfeldTkatchenko2020} surrogate model (see computational details) results in a much higher leave-one-out prediction error of 10.4 kcal/mol.
By comparison, generic QM methods such as semi-empirical PM7 method\cite{PM7}, 
density functional theory (DFT/PBE0/def2-TZVP),
and coupled cluster single double perturbative triples F12 calculations, (taken from  G2 Ref.~\cite{G2})
produce MAEs of 9.5, 1.2, and 0.6\,kcal/mol, respectively.
This indicates that our model can achieve accurate descriptions of trends in chemical compound space.
We should caution, however, that our model is biased due to the calibration, and is likely to perform significantly worse for other chemistries.

%TODO move parentheses and CV to methods

\subsection{Comparison to Pauling}
Pauling's model and our model are both simple expression with similar overall accuracy.
Pauling's bond model \cite{pauling} expresses the BDE as
\begin{equation}\label{eq:pauling}
    E_\text{AB} = \frac{E_\text{AA} + E_\text{BB}}{2} + 23 (\chi_\text{A} - \chi_\text{B})^2
\end{equation}
which depends on the the homolytic binding energies $E_\text{AA}, E_\text{BB}$
and the electronegativites $\chi_\text{A}, \chi_\text{B}$.
The model is based on a wavefunction decomposition into a covalent and an ionic part. Pauling proposed that the bond energy could accordingly be split into a covalent contribution, approximated as the mean of the homolytic bond energies, and an ionic contribution, postulated to be represented by the difference in electronegativies $(\chi_\text{A}-\chi_\text{B})^2$. 
Pauling optimized electronegativities
in order to reproduce binding energies as accurately as possible.\cite{pauling}
This model's predictions for our test sets were discussed before  (See also FIGs. \ref{fig:bde}, \ref{fig:compare_models}).

% Paulings electronegativities are fitted to experimental bond enthalpies not PBE0 energies, should we refit them to our model to get a fair comparison?
Note that Paulings expression requires knowledge of the homolytic bond formation energies and introduces electronegativities as additional quantity while our model relies directly on the nuclear charges.
While electronegativity is useful to explain trends in chemical properties, our formulation depending directly on nuclear charges is directly connected to the external potential in the electronic Hamiltonian, and thereby more rigorously rooted in the fundamental physics governing chemistry.
Furthermore, homo-diatomics in the third and fourth period dominate the error of our model, while these cases are direct model parameters for Pauling. 
%(P-P, S-S, As-As, Se-Se). 
% Why? is this maybe an error of DFT or are there some different/additional interaction taking place? check with NIST values if possible
%Paulings model recovers the exact energy in these cases because it requires the BDE of the homo-diatomics as a variable. 
%TODO: refit Pauling, if trend electronegativity persists, comment on that
Note that our model outperforms Pauling's if the binding partners A and B have a large electronegativity difference as observable for example for the C-F, Si-Cl or N-F bond (FIG.~\ref{fig:bde}).

\subsection{Comparison to Hammett's equation}
Another empirical model to quantify property trends across chemical spaces was proposed by Hammett more than 80 years ago~\cite{hammett35, hammett37}, 
\begin{equation}
    P \approx \rho \,\cdot\,\sigma.
\end{equation}
Originally, $P = \log[K/K_0]$ was the equilibrium constant for various reactions of benzene derivatives normalized with respect to a reference reaction constant $K_0$, $\sigma$ described the effect of different substituents and $\rho$ accounted for the reaction type (e.g. mechanism or solvent).
However, the model has been used to describe many other properties like activation energies\cite{Bragato2020}, orbital energies of metal organic complexes\cite{C9SC02339A} or dipole moments\cite{Hammett_dipole}. Furthermore, the model has been applied to non-benzyl compounds.\cite{Bragato2020,Hammett_silyl}
The relation between Hammett constants and electronegativity has also been noted in the context of nucleo- and electro-philicity relevant for mechanistic discussions in organic chemistry.~\cite{hammett_en}
While Hammett's model is very intuitive since it only requires separability of two dominating variables, its physical motivation has remained unclear.~\cite{frenking_hammett}

\begin{figure}[ht]
\centering
\includegraphics[width=0.4\textwidth]{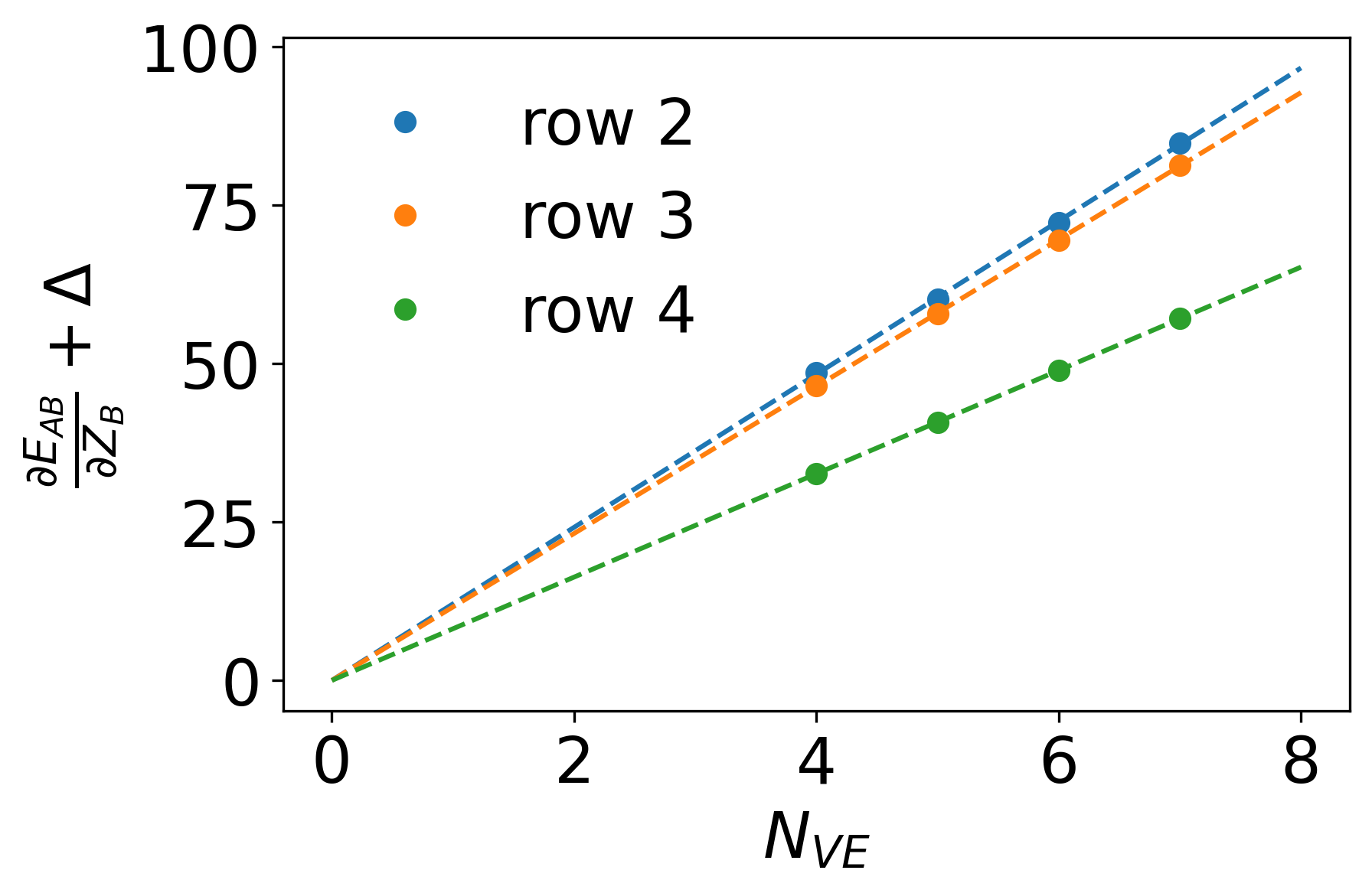}
\caption{The sensitivity of the binding energy due to change in composition expressed by a Hammett type relationship between the derivative of the binding energy with respect to the nuclear charge $\pdv{E_\text{AB}}{Z_\text{B}}$ and the number of valence electrons $N_\text{VE}$ of binding partner B for binding in different rows of the periodic table. The curves are shifted by a constant $\Delta$ such that they intersect the origin.}
\label{fig:hammett}
\end{figure}

Our model might offer a rationalization of Hammett's because it emerges from the quantum alchemy based atomic energy decomposition and accounts for the change in binding energy with respect to composition, fully consistent with Hammett's approach.
In particular, the change of binding energy with respect to the nuclear charge, the partial derivative of our model, is simply given by
\begin{equation}\label{eq:ABF}
    {\pdv{E_\text{AB}}{Z_\text{B}}} (Z_\text{A}, Z_\text{B})  = -b Z_\text{A} + \frac{7}{3} c Z_\text{B}^{4/3}.
\end{equation}
FIG.~\ref{fig:hammett} displays the near-linear trend of this derivative for a fixed value of $Z_\text{A}$ as a function of the number of valence electrons $N_\text{VE}$ of $Z_\text{B}$ for each period investigated in this study. 

We have shifted each curve by a constant $\Delta$ such that it intersects the origin at $N_\text{VE} = 0$. The curves are approximately linear in the number of valence electrons with the slopes being proportional to $\frac{7}{3} c$ (see Eq.~\eqref{eq:ABF}). 
Thus, ${\pdv{E_\text{AB}}{Z_\text{B}}} + \Delta$ can be modelled by a Hammett ansatz with $\sigma = N_\text{VE}$ and $\rho$ accounting for different binding behaviour due to a change in number of core electrons for different rows.
As the row number increases the slope becomes flatter indicating a lower sensitivity of the binding energy to a change of the binding partner as observable in FIG.~\ref{fig:bde}. The drastic decrease of the slope from the third to the fourth row could be due to the additional 10 3\emph{d} core electrons for elements in the fourth row.

The identification of the number of valence electrons of binding partner B as the $\sigma$-parameter is possible because the influence of the binding partner B is expressed as a function of the nuclear charge in our model. We have arrived at this expression based on the atomic energy decomposition within quantum alchemy. We believe that these findings indicate that such decomposition into atomic or fragment contributions can deepen our understanding of empirical rules such as Hammett's model. Note that this finding is also consistent with multiple other studies which found a correlation of the $\sigma$ parameters with atomic quantities like NMR-shifts~\cite{hammett_NMR1, hammett_NMR2}, polarizing force\cite{hammett_polarforce}, fragment self similarity measures\cite{hammett_self_simi} or atomic charges\cite{hammett_atomic_ch1,hammett_atomic_ch2}.

\subsection{Limitations to covalent bonds}
\begin{figure}[ht]
\centering
\includegraphics[width=0.4\textwidth]{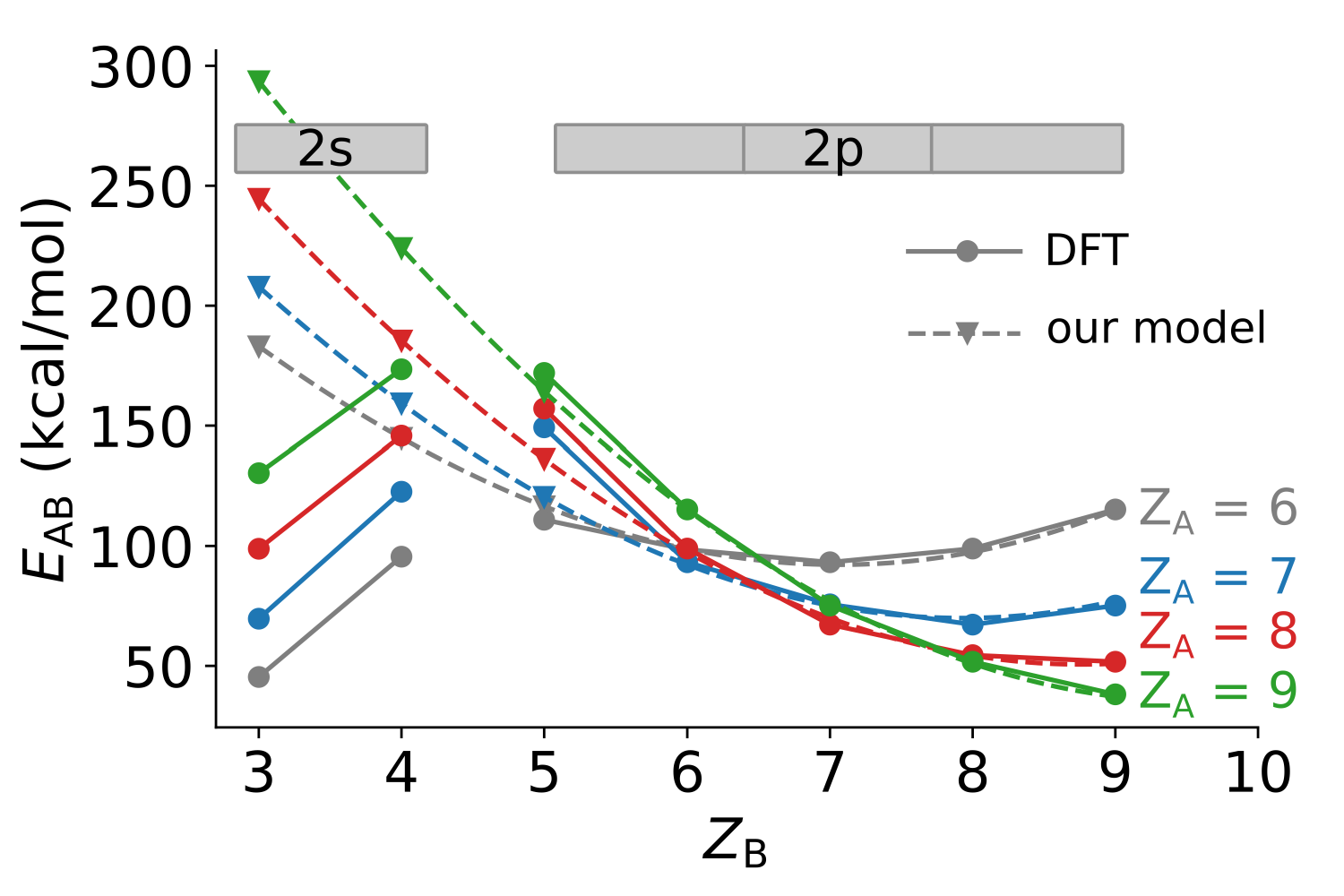}
\caption{Limitations of applicability: BDEs between atom A and B corresponding to C/N/O/F or
Li/Be/B/C/N/O/F, respectively.
For second quantum number differing between A and B, our model [Eq.~\eqref{eq:SRL*}] breaks down.}
\label{fig:mg_trends}
\end{figure}

So far, we have only considered typical covalent bonding scenarios between $p$-block elements, i.e.~elements from groups IV-VII in the periodic table. 
These elements all have in common that their valence electrons share the same second angular momentum quantum number. 
Our model has not been developed for decreased covalent character, i.e.~for bonding atoms with differing second quantum numbers.
Not surprisingly, for example, BDEs for single bonds shown in  
FIG. \ref{fig:mg_trends}
indicate a qualitatively different behavior when one bonding partner, 
say atom B, comes from the alkaline or earth-alkaline group. 
According to our model calibrated for $p$ elements only, bond energy content would monotonically increase as the nuclear charge $Z_B$ decreases.
In reality, however, bond energy content must decrease as B changes from predominantly covalent bonding among $p$-block elements into ionic bonding regimes with FF and LiF as the two opposite extremes.

\subsection{Bond order}
Close inspection of results for the $p$-block elements shown in FIG. \ref{fig:mg_trends} indicates that our model is the least accurate when it comes to the prediction of BDEs  involving Boron ($Z_B = 5$).
This could be due to the parameters being optimized for bonds with bond orders being close to 1. 
B-N, B-O and B-F bonds, however, are known to have a bond order
of approximately 1.4.\cite{mayer1,mayer2,mayer3}. 
Consequently, one should expect our covalent single bond model to systematically underestimate the binding energies for these systems --- consistent with the numerical observation.

% double bond old version
% We have investigated whether our model can be re-calibrated to also predict binding energies for other bond orders. More specifically, we have refitted parameters $a, b, c$  to the binding energies of double bonds A=B for all six possible combinations of carbon, nitrogen and oxygen, saturated with hydrogen.
% The resulting model also accounts for double bond energies (MAE $\sim$ 0.9~kcal/mol) as long as all reference systems are in the same electronic singlet state. 
% % As shown in SI~FIG.~\ref{fig:bde_double}, the inclusion of 
% % O$_2$ in its triplet ground-state worsen.
% Using the binding energy of the O$_2$ triplet ground state instead, not surprisingly the predictions worsen, in this case such that the MAE becomes 7.4~kcal/mol (SI~FIG.~\ref{fig:bde_double}).

To further explore this aspect, we have investigated whether our model can be re-calibrated to predict binding energies for other bond orders. We have considered double bonds A=B for all six possible combinations of elements from main groups IV, V and VI within the same row of the periodic table, e.g. carbon, nitrogen and oxygen for the second row. In contrast to the single bonded molecules, the spin state is not the same for all double bonded systems. O$_2$, S$_2$ and Se$_2$ have triplet ground states while the other molecules are singlets. To remove the additional complexity due to a change in spin state, we used the same singlet spin state for all double bonded molecules, i.e. we refitted the parameters $a, b, c$ to energy differences $\Delta E^\text{TS}$ between fragments in a triplet state after homolytic bond cleavage, and the bound molecule in a singlet state. The trend in these binding energies is very similar to the one for single bonds with a decrease in binding energy as the nuclear charges of the binding atom grows. 
The resulting model's performance is equally good with prediction errors of 0.9, 1.3 and 2.0\,kcal/mol for the second, third and fourth row, respectively (see SI with Fig.~S1, and Tab.~S1 for optimized parameters).
%(see SI with Fig. \ref{fig:bde_double}, and TABLE~\ref{tab:coeff_TS} for optimized parameters).
\subsection{Electronic configuration}
To further investigate the impact of spin, we re-calibrated our model to energy differences $\Delta E^\text{GS}$ between the ground states of fragments and molecules, e.g. O$_2$ in a triplet instead of a singlet state (Fig.~S2 A, Tab.~S2) and the energy differences $\Delta E^\text{TT}$ and $\Delta E^\text{SS}$ with fragments and molecules either both in triplet (Fig.~S2 B, Tab.~S3) or both in singlet states (Fig.~S2 C, Tab.~S4).
The prediction of $\Delta E^\text{GS}$ yields a MAE of 7.4\,kcal/mol for compounds in the second row. This increase in error with respect to $\Delta E^\text{TS}$ could be attributed to mixed spin states of the molecules. However, for the third and fourth row prediction errors for $\Delta E^\text{GS}$ are 1.6 and 1.9\,kcal/mol, which is similar to the error for $\Delta E^\text{TS}$. For these rows, the spin state changes not only for the molecules but also for the fragments. The fourth main group fragments (SiH$_2$, GeH$_2$) have a singlet ground state while the other fragments are triplets. Thus, changing spin states do not generally lower the accuracy.

The prediction errors for $\Delta E^\text{TT}$ and $\Delta E^\text{SS}$ are around 7-8\,kcal/mol for the second row and 10-14\,kcal/mol for the third and fourth row (see Tab.~S3 and Tab.~S4 for details). The worse performance compared to single bonded systems is not surprising since the functional form of our model is inspired by the alchemical decompositioning of bond dissociation energy trends for single bonds in the ground state, that are significantly different from $\Delta E^\text{TT}$ and $\Delta E^\text{SS}$.

Finally, we have also recalibrated our model (Eq.~\eqref{eq:SRL*}) to fit randomly drawn points from an uniform distribution in order to assess in how far the performance of our model can be attributed to its mathematical flexibility or rather to its inherent capability to account for the underlying physics of the studied systems (see Fig.~S3). The MAE of the randomly drawn data is in all cases, except for $\Delta E^\text{TT}$ for the fourth row, substantially higher than for the binding energy differences. This finding corroborates the notion that our model's performance is not coincidence but rather due to its appropriate functional form approximating the relevant physics to a certain degree.

\subsection{Trends and dependence on period}
Since many questions in chemistry only require knowledge about differences in BDEs, we have investigated the applicability of our model towards the prediction of trends among bonds. 
More specifically, 
generalization of our model to deal with covalent bonding within {\em any} row from the $p$-block with just 3 parameters $a, b,c $ is impossible due to the large differences in nuclear charges with increasing row number $n$. We find, however, that $a, b,$ and $c$ vary  smoothly with $n$ when predicting changes in BDEs, i.e.~ $\Delta E = E_\text{AB} - E_\text{AC}$,  in combination with a second order Taylor expansion of the $Z^{7/3}$-terms. 
Then, parameters can be described as simple functions of the principal quantum number $n$ (the row number) and $\Delta E$ is approximately given by
\begin{equation}\label{eq:SRL*_diff}
    \Delta E  \approx \Delta Z \left[28 (n-1) + (8.5 + (n-3)^2) Z_\text{A} \right] + 6 (Z_\text{B}^2 - Z_\text{C}^2)
\end{equation}
with the change in nuclear charges $\Delta Z = Z_\text{C} - Z_\text{B}$ (see SI for details).

\begin{figure}[ht]
\centering
\includegraphics[width=0.4\textwidth]{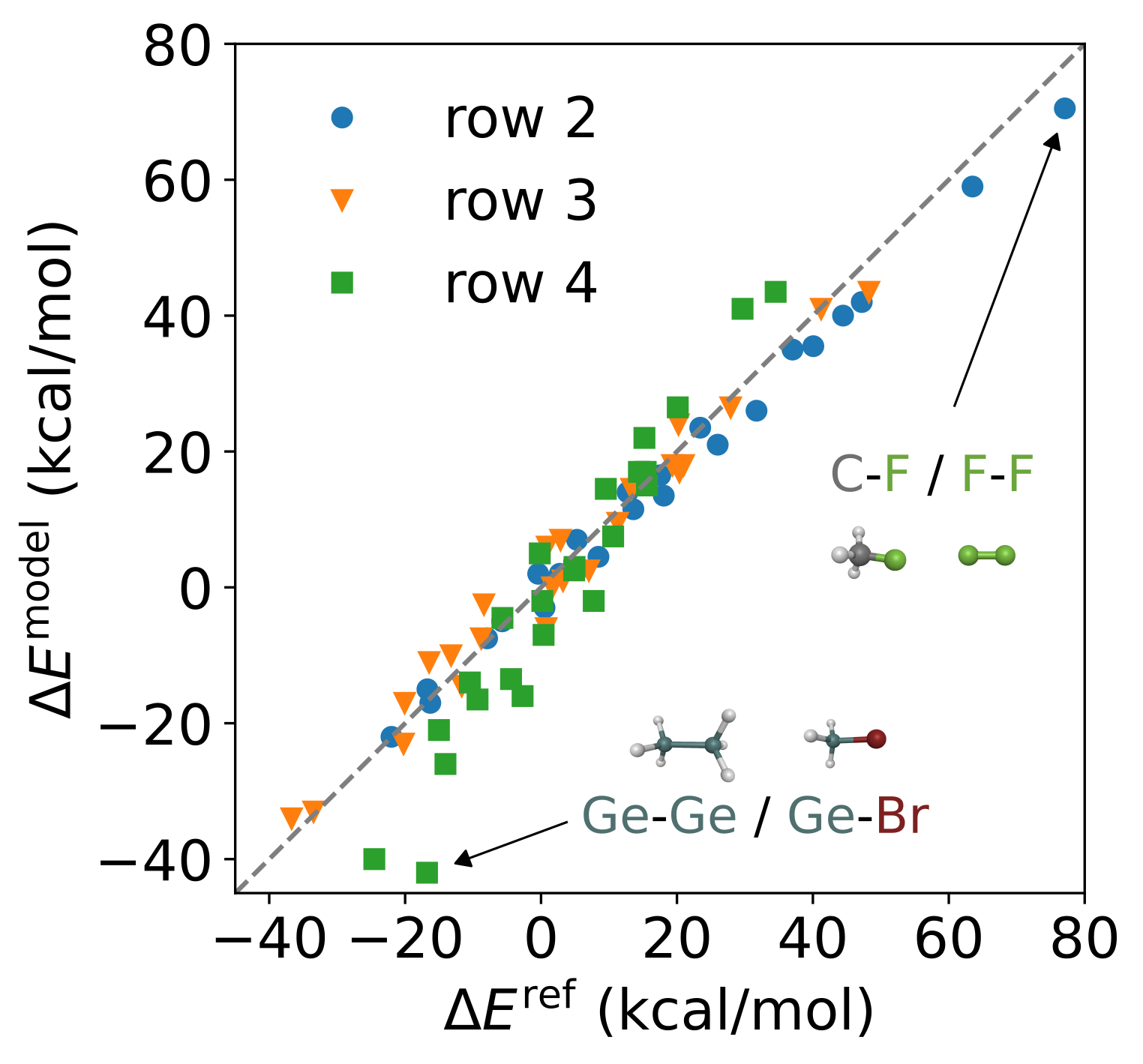}
\caption{Trends in covalent binding: Predicted (Eq.~\eqref{eq:SRL*_diff}) vs.~Truth (corresponding to PBE0/def2-TZVP). 
72 binding energy differences between bonds A-B and A-C ($\Delta E$),
both within either the second, third, or fourth row of the periodic table, $\Delta E = E_\text{AB} - E_\text{AC}$, where $Z_\text{B} \le Z_\text{C}$.
Corresponding molecules involved are on display in FIG.~2.
}
\label{fig:approx_eq}
\end{figure}

Encouragingly, Eq.~\ref{eq:SRL*_diff} meaningfully reproduces changes in covalent bond energies covering ranges from -40 to +70 kcal/mol with a MAE of just 4.2\,kcal/mol with respect to density functional theory.
Results are shown in FIG.~\ref{fig:approx_eq}
for any single bond changes within second, third, or fourth row of the periodic table. 

Albeit of interest, the generalization of Eq.~\ref{eq:SRL*_diff} to also account for atoms A and B and C coming from {\em different} rows, or to involve other bond-orders, has not yet been explored in this study. 

% We also tried to generalize our model to describe bonding in multiple rows with a single set of parameters. The abrupt increase in nuclear charge with increasing row number poses a challenge. Therefore, we focused on prediction of binding energy differences $\Delta E = E_\text{AB} - E_\text{AC}$ of bonds A-B and A-C because this eliminates the strongly row dependent variable $a$ from the binding expression. Furthermore, the $Z^{7/3}$ terms are approximated by a second order Taylor expansion.
% This allows us to express $\Delta E$ mainly as a function nuclear charge differences that are row independent resulting in a smoother change of the parameters from row to row. 
% leading to 
% \begin{equation}
% \begin{split}
%     \Delta E_\text{ABC} &= -28 (n-1) \Delta Z_\text{BC} - ([n-3]^2 - 3.5) Z_\text{A} \Delta Z_\text{BC} \\
%     &+ 6 \left( \Delta Z_\text{AB}^2 - \Delta Z_\text{AC}^2 \right).
% \end{split}
% \end{equation}

% the model works well for isoelectronic systems/from p-block with same spin configuration

\section{Conclusion}
We have presented a simple and, to the best of our knowledge, novel expression for covalent binding energies in terms of nuclear charges. Despite its simplicity, it is deeply rooted in the underlying physics of quantum mechanics via the computational alchemy based reasoning.  The expression might prove useful for developing an improved intuition regarding trends 
of binding energies  across chemical compound space. It has only three calibration parameters which can easily be regressed to available reference data. 

We have found the model to be limited to covalent single and double bonds among
atoms with the same first and second quantum number. We have presented promising numerical evidence for $p$-block atoms (except for rare gas elements) coming from 2nd, 3rd, and 4th row. We have compared our model to Pauling's electronegativity model, and we have discussed its consistency with respect to Hammett's equation. 
We note that the description of ionic, metallic, or van der Waals bonding through inclusion of $s$ and $d$ block elements is still outstanding and will be part of future research. This also holds for the generalization to bonds involving elements that differ in principal quantum numbers.
Interactions beyond the two binding partners, 
e.g. the influence of immediate substituents via inductive or mesomeric effects, should be studied for further generalization.
But also the inclusion of other environmental effects such as aromaticity, 
van der Waals interactions (hydrogen bonding, London dispersion), 
or electric field effects (from static multipole moments or externally) could extend the applicability of the model.
% For further generalization of the model, interactions beyond the two binding partners should be considered.
% In particular, the influence of immediate substituents, e.g. via inductive or mesomeric effects, should be studied.
% Also the inclusion of other environmental effects such as aromaticity, van der Waals interactions (hydrogen bonding, London dispersion), 
% or electric field effects (from static multipole moments or externally) could extend the applicability of the model.

Conceptionally speaking, our model relies on coarse-graining the expectation value of the electronic Hamiltonain throughout chemical space. 
As such, it is consistent with quantum mechanics and offers a fresh perspective on bond dissociation energies which is in line with Hammett's expression.
Historically speaking, it represents an equally powerful yet possibly less empirical alternative to Pauling's electronegativity model.
Future work will show to which extent this partitioning approach can be used to deepen our understanding of chemical space with respect to other extensive properties, and if it is useful for computational materials and molecular design efforts.

% -improved intuition/understanding helps with more data-efficient physics-inspired QML models (which reduces data-needs ~ CO2-footprint ...
% -future: Explore simple models of other properties than bonds ...
\section{Methods}
\subsection{Construction of the model}
\begin{figure}[!ht]
\centering
\includegraphics[width=0.3\textwidth]{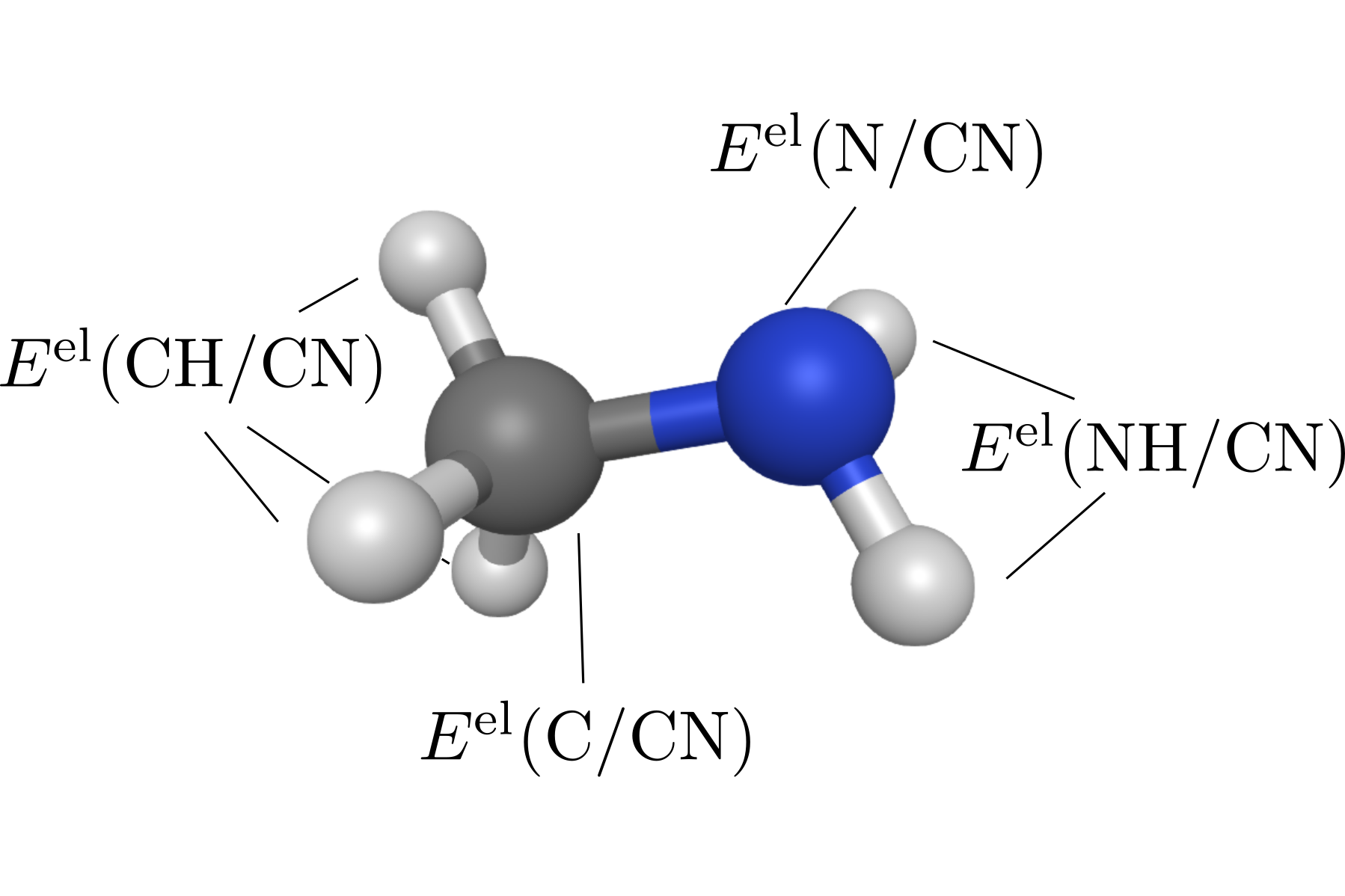}
\caption{
Atomic electronic energy decomposition notation exemplified for methylamine.
}
\label{fig:aed}
\end{figure}

% Continuing from Eq.~\eqref{eq:atomic_energy}, 
% the electronic energy $E_\text{AB}^\text{el}$ of a saturated diatomic A-B
% can be decomposed into atomic contributions (see FIG.~\ref{fig:aed} exemplary for methylamine)
% \begin{equation}
%     E_\text{AB}^\text{el} = E_\text{A}^\text{el}(\text{AB}) + E_\text{B}^\text{el}(\text{AB}) + E_\text{AH}^\text{el}(\text{AB}) + E_\text{BH}^\text{el}(\text{AB}),
% \end{equation}
% where $E_\text{A}^\text{el}(\text{AB})$ and $E_\text{B}^\text{el}(\text{AB})$ are the atomic energies of A and B and $E_\text{AH}^\text{el}(\text{AB})$ and $E_\text{BH}^\text{el}(\text{AB})$ are the atomic energies of the hydrogens attached to A and B, respectively.
Continuing from Eq.~\eqref{eq:atomic_energy}, 
the electronic energy $E^\text{el} (\text{AB})$ of a saturated diatomic A-B
can be decomposed into 
\begin{equation}
\begin{split}
E^\text{el}(\text{AB}) &= E^\text{el}(\text{A/AB}) + E^\text{el}(\text{B/AB}) \\
&+ E^\text{el}(\text{AH/AB}) +E^\text{el}(\text{BH/AB})
\end{split}
\end{equation}
where $E^\text{el}(\text{A/AB})$ and $E^\text{el}(\text{B/AB})$ are the atomic energies of A and B and $E^\text{el}(\text{AH/AB})$ and $E^\text{el}(\text{AH/AB})$ are the sum of atomic energies of the hydrogens attached to A and B, respectively (see Fig.~\ref{fig:aed} for an example).
% would a sketch of a saturated diatomic be helpful

Furthermore, the electronic contribution to the binding energy $E_\text{AB}^\text{el}$ is the difference between the electronic energies of the fragments A and B generated by homolytic bond cleavage and the electronic energy of compound AB
\begin{equation}
    E_\text{AB}^\text{el} = E^\text{el}(\text{A}) + E^\text{el}(\text{B}) - E^\text{el}(\text{AB}).
\end{equation}

The electronic contribution to the \emph{atomic} binding energy, e.g. $E_\text{AB}^\text{el} (\text{A})$ for atom A, is then defined as
\begin{equation}\label{eq:atomic_electronic_binding:energy}
    E_\text{AB}^\text{el} (\text{A}) = E^\text{el}(\text{A/A}) - E^\text{el}(\text{A/AB})
\end{equation}
where $E^\text{el}(\text{A/A})$ and $E^\text{el}(\text{A/AB})$ are the atomic energies of A in fragment A and diatomic compound AB, respectiveley.

% One can decompose the electronic binding energy into its atomic contributions, $ E^{\text{el}}_\text{AB} = E_\text{A}^\text{el} + E_\text{B}^\text{el}$, with  A and B %are elements of second row and fourth to seventh main group of the periodic table of elements.
% corresponding to saturated elements carbon, nitrogen, oxygen or fluorine.
Calculating such atomic binding energies according to APDFT~\cite{VonRudorff2019},
we have observed that $E_\text{AB}^\text{el} (\text{A})$ increases approximately linearly with the nuclear charge $Z_\text{B}$ of binding partner B for saturated diatomics composed from the elements carbon, nitrogen, oxygen and fluorine (Fig.~S4 A). Furthermore, the atomic energy of the hydrogen atoms remains approximately constant with varying binding partner (Fig.~S4 B).

% \begin{equation}
%     E_\text{A}^\text{el} \approx m_\text{A} Z_\text{B} + n_\text{A}.
% \end{equation}
This enables us to express the binding energy approximately as
\begin{equation}\label{eq:SRL_el}
E_\text{AB}^\text{el} \approx
 \underbrace{\beta_\text{A} Z_\text{B} + \alpha_\text{A}}_{=E_\text{AB}^\text{el} (\text{A}) + E_\text{AB}^\text{el} (\text{AH})} + \underbrace{\beta_\text{B} Z_\text{A} + \alpha_\text{B}}_{=E_\text{AB}^\text{el} (\text{B}) + E_\text{AB}^\text{el} (\text{BH})}.
\end{equation}
The energy contributions of fragments A and B are characterized by the parameters $\alpha_\text{A}$, $\beta_{A}$ and $\alpha_\text{B}$, $\beta_{B}$, respectively. $\alpha$ accounts for the constant contribution of the heavy atom (A or B) and of the hydrogens attached to it to the binding energy, while $\beta$ describes the contribution of the heavy atom for varying binding partners.
Furthermore, we approximate the contribution of the nuclear repulsion to the binding energy $E_\text{AB}^\text{nuc}$ by only considering the interaction between the heavy atoms A and B at an average bond distance $\bar{d}$. This average is calculated from the bond lengths between A and B for all considered compounds A-B.
Consequently, the total binding energy $E_\text{AB}$ including nuclear repulsion can be expressed as
\begin{equation}\label{eq:SRL}
E_\text{AB} \approx
\underbrace{\beta_\text{A} Z_\text{B} + \alpha_\text{A}}_{=E_\text{AB}^\text{el} (\text{A}) + E_\text{AB}^\text{el} (\text{AH})} + \underbrace{\beta_\text{B} Z_\text{A} + \alpha_\text{B}}_{=E_\text{AB}^\text{el} (\text{B}) + E_\text{AB}^\text{el} (\text{BH})}
 - \underbrace{\frac{Z_\text{A} Z_\text{B}}{\bar{d}}}_{=E_\text{AB}^\text{nuc}}.
\end{equation}

% we do not give an explanation why we can approximate the nuclear repulsion this way
% the motivation is that nuclear repulsion from hydrogens is independent of binding partner more detail on this?

% why don't we use atomic energies?
The parameters $\alpha, \beta$ are determined from a least squares fit to $E_\text{AB} + \frac{Z_\text{A}Z_\text{B}}{\bar{d}}$ for all ten unique combinations {A-B} of carbon, nitrogen, oxygen and fluorine. The binding energies $E_\text{AB}$ are calculated with DFT (see Computational Details).

% Since the atomic energies obtained from Eq.~\eqref{eq:atomic_energy} depend on a numerically inaccurate integration scheme, we obtain the parameters $\alpha, \beta$ from a fit to binding energies calculated with DFT (see Computational Details). This is possible because Eq.~\eqref{eq:SRL} does not explicitly depend on the atomic energies anymore. 

While Eq. \eqref{eq:SRL} is based on trends for diatomics in the second row of the periodic table, we also apply it to diatomics from the fourth to seventh main group of the third and fourth row.
% Fig. \ref{fig:bde} shows DFT reference binding energies (DFT, \includegraphics[height=\fontcharht\font`\b]{./figures/ref_label.png}) for the saturated diatomics of the fourth to seventh main group and second to fourth row of the periodic table.
% The binding energies of A-B decrease with increasing nuclear charge of atom A for the diatomics in the second row. Binding energies in the third and fourth row follow a similar trend because the valence electron configuration is the same across the different rows. Due to this similarity, Eq. \eqref{eq:SRL} should also be applicable to the third and fourth row of the periodic table.
% The binding energies predicted by Eq. \eqref{eq:SRL} with optimized parameters are shown in Fig. \ref{fig:bde} (SRL\includegraphics[height=\fontcharht\font`\b]{./figures/SRL_label.png}).

The mean absolute errors (MAEs) for binding energy predictions with Eq.~\eqref{eq:SRL} are 0.5, 0.4 and 0.3\,kcal/mol for rows 2-4, respectively and the optimized parameters can be found in Tab.~S5. The model accurately reproduces the binding energies and can be applied to different rows of the periodic table. 
However, it is also prone to overfitting because it uses eight parameters to model binding energies for ten compounds. Thus, we studied the relation of the parameters $\alpha, \beta$ on the nuclear charge of the respective element in an attempt to reduce the number of parameters.
% While the models have been developed from atomic energy trends for compounds of the second row of the periodic table, the predictions are highly accurate across all rows with a MAE of 0.4\,kcal/mol for Eq. \eqref{eq:SRL} and a MAE of 1.8\,kcal/mol for Eq. \eqref{eq:SRL*}.
% The higher accuracy of Eq. \eqref{eq:SRL} compared to Eq. \eqref{eq:SRL*} is due to the higher number of parameters (8 vs 3) giving a higher flexibility to the model but making it also more prone to overfitting.
% Thus, we will focus on Eq. \eqref{eq:SRL*} in the following.

The optimized parameter $\beta_\text{A}$ correlates linearly with $Z_\text{A}$
\begin{equation}\label{eq:beta}
    \beta_\text{A} \approx b' Z_\text{A}.
\end{equation}
as shown in Fig.~S5 A. 
Thus, the terms $\beta_{A} Z_\text{B}$ and $\beta_{B} Z_\text{A}$ in Eq. \eqref{eq:SRL} can be written as
\begin{equation}
    \beta_\text{A} Z_\text{B} + \beta_\text{B} Z_\text{A} \approx 2 b' Z_\text{A} Z_\text{B}.
\end{equation}
The $\beta$-terms account mainly for the large nuclear repulsion term $Z_\text{A}Z_\text{B}\bar{d}^{-1} \gg E_\text{AB}$ in the dependent variable $E_\text{AB} + \frac{Z_\text{A}Z_\text{B}}{\bar{d}}$.
We note that the linear dependence of $\beta_\text{A}$ on $Z_\text{A}$ is also consistent with
the relation 
\begin{equation}\label{eq:alch_enan}
    E^\text{el}_\text{QR} \approx E^\text{el}_\text{SR} + \frac{1}{2} \left( E^\text{el}_\text{QQ} + E^\text{el}_\text{SS} \right),
\end{equation}
between binding energies of elements Q, R and S with $Z_\text{Q} = Z_\text{R} - 1 = Z_\text{S} - 2$, that was derived from alchemical enantiomers.\cite{VonRudorff2021} Insertion of the definition for electronic binding energies from Eq.~\eqref{eq:SRL_el}
into Eq.~\eqref{eq:alch_enan} yields
\begin{equation}
    \begin{split}
        \beta_\text{Q} Z_\text{R} + \alpha_\text{Q} + \beta_\text{R} Z_\text{Q} + \alpha_\text{R} &= \beta_\text{S} Z_\text{R} + \alpha_\text{S} + \beta_\text{R} Z_\text{S} + \alpha_\text{R}\\ &+  \beta_\text{Q} Z_\text{Q} + \alpha_\text{Q} - \beta_\text{S} Z_\text{S} - \alpha_\text{S} \\
        \beta_\text{Q} Z_\text{R} + \beta_\text{R} Z_\text{Q} &= \beta_\text{S} Z_\text{R} + \beta_\text{R} Z_\text{S} + \beta_\text{Q} Z_\text{Q} - \beta_\text{S} Z_\text{S}\\
        \beta_\text{R} \underbrace{(Z_\text{Q} - Z_\text{S})}_{= -2} &= \beta_\text{S} \underbrace{(Z_\text{R} - Z_\text{S})}_{= -1} + \beta_\text{Q} \underbrace{(Z_\text{Q} - Z_\text{R})}_{= -1}\\
        \beta_\text{R} &= \frac{\beta_\text{S} + \beta_\text{Q}}{2},
    \end{split}
\end{equation}
which implies a linear relation between the different values for $\beta$.
% The Eq. \eqref{eq:alch_enan} is valid if the nuclear charges of Q, R, S are
% \begin{equation}\label{eq:Z_rel}
%     Z_S = Z_R + 1 = Z_Q + 2.
% \end{equation}
% Since only elements from main group IV to VII within one row are considered, the condition Eq. \eqref{eq:Z_rel} is approximately fulfilled.

The optimized offset $\alpha_\text{A}$ has a non-linear relationship with $Z_\text{A}$ (Fig.~S5 B).
Thus, we model the dependence of $\alpha_\text{A}$ on $Z_\text{A}$ as
\begin{equation}\label{eq:alpha}
    \alpha_\text{A} \approx c Z_\text{A}^{\gamma} + a',
\end{equation}
where the exponent $\gamma$ accounts for the non-linearity.

Substitution of $\beta$ and $\alpha$ in Eq. \eqref{eq:SRL} with the expressions in Eq. \eqref{eq:beta} and Eq. \eqref{eq:alpha} and rearrangement of the resulting equation yields
\begin{equation}\label{eq:SRL_gamma}
    E_\text{AB} \approx a - b Z_\text{A} Z_\text{B} + c\left( Z_\text{A}^{\gamma} + Z_\text{B}^{\gamma} \right)
\end{equation}
with $a = 2 a'$ and $b = \left( \frac{1}{\bar{d}} - 2 b' \right)$.
% The optimization of $a, b,c, \gamma$ with respect to $E_\text{AB} + \frac{Z_\text{A}Z_\text{B}}{\bar{d}}$ for each row independently reveals that the optimal exponents $\gamma_\text{opt}$ are similar for the different rows (see SI~Fig.~\ref{fig:gamma}). Thus, $\gamma$ was kept the same for all rows and only $a, b, c$ were optimized for each row independently. The optimal value for $\gamma$ in this optimization procedure was $\gamma_\text{opt} = 2.31 \approx 
% \frac{7}{3}$.
Optimization of the parameters in Eq.~\eqref{eq:SRL_gamma} by a combination of non-linear least squares for $a,b,c$ and a line scan for $\gamma$ leads to similar values for the optimal exponent $\gamma_\text{opt}$ for the different rows (see Fig.~S6). Thus, $\gamma$ is kept the same for all rows and only $a, b, c$ are optimized for each row independently. The optimal value for $\gamma$ in this optimization procedure is $\gamma_\text{opt} = 2.36 \approx 
\frac{7}{3}$.
%that the optimal exponent $\gamma$ is very similar for the different rows ($\gamma_2 = 2.41, \gamma_3 = 2.18, \gamma_4 = 2.56$) and close to . Thus,
% only the parameters $a,b,c$ were optimized for each row individually while $\gamma$ was fixed to be the same for each row. The optimal value for $\gamma$ in this optimization procedure was $\gamma = 2.31$.
This is an interesting result because $-0.768745 \cdot Z^{7/3}$ is the leading term in an expansion of the energy of a free atom in its nuclear charge.~\cite{Scott, Schwinger}
Hence, the binding energy can be expressed as
\begin{equation}
\begin{split}
    E_\text{AB} &\approx a - b Z_\text{A} Z_\text{B} + c\left( Z_\text{A}^{7/3} + Z_\text{B}^{7/3} \right) \\ &\approx a - b Z_\text{A} Z_\text{B} + c'\left( E_\text{A}^\text{atom} + E_\text{B}^\text{atom} \right)
\end{split}
\end{equation}
with the energies of the free atoms $E_\text{A}^\text{atom}$ and $E_\text{B}^\text{atom}$ and $c' = -\frac{c}{0.768745} $.
% The optimized parameters $a,b,c$ with $\gamma = 7/3$ and respective MAEs are shown in Table \ref{tab:coeff} for all rows. The predicted binding energies are presented in Fig.~\ref{fig:bde} (our model).

\section{Computational Details}
The atomic energies (Eq.~\eqref{eq:atomic_energy}) of the saturated diatomics were calculated using geometries from the amons dataset\cite{Huang2020}. The fragment structures were generated by splitting the homo-diatomics without further geometry optimization. 
The required electron densities were obtained following the procedure in earlier work\cite{VonRudorff2019} from
calculations with the CPMD\cite{CPMD} code and the CPMD2CUBE program\cite{CPMD2CUBE} using the PBE\cite{pbe} functional in a plane wave basis with a cutoff of 200~Ryd, GTH\cite{Goedecker1, Goedecker2} pseudopotentials and a wavefunction gradient convergence set to $10^{-6}$ making partially use of gnu parallel.\cite{Tange2011a} A primitive cell with a box length of 14.338\,\AA \ was used for saturated diatomics and of 11.380\,\AA \ for the fragments. The GTH pseudopotential parameters were scaled by $\lambda = 6/14, 8/14, 11/14, 1$ and $\lambda = 3/7, 4/7, 6/7, 1$ for the saturated diatomics and the fragments, respectively to generate electron densities for different values of $\lambda$. The electron density at $\lambda = 0$ was represented as a uniform distribution.
The integration with respect to $\vec{r}$ and $\lambda$ was carried out as weighted summation over grid points and with the trapezoidal rule, respectively. 
%For additional information see previous work by Rudorff and von Lilienfeld.

The binding energies to determine the optimal parameters in Eq.~\eqref{eq:SRL*} and \eqref{eq:SRL} were calculated with PySCF\cite{pyscf, pyscf2, pyscf3, pyscf4} for single bonded systems and with GAUSSIAN\cite{Gaussian09} for double bonded systems with PBE0/def2-TZVP\cite{def2} (restricted open shell for fragments). Reported binding energies are for optimized geometries of saturated diatomics and fragments.
Initial guesses for the structures were generated through the LERULI API\cite{leruli, rdkit, xtb1, xtb2}.
The parameter optimization in Eq.~\eqref{eq:SRL*}, \eqref{eq:SRL*_diff}, \eqref{eq:SRL}, \eqref{eq:SRL_gamma} was performed by linear or non-linear least squares fitting as implemented in numpy\cite{numpy} and scipy\cite{scipy}.
Binding energies of our model presented in Fig.~\ref{fig:compare_models} are with respect to W4-17 data after determining the parameters via leave-one-out crossvalidation.
The machine learning predictions were obtained from kernel ridge regression with a Gaussian kernel and the bag of bonds\cite{bob} representation in leave-one-out crossvalidation. PM7 binding energies were calculated with MOPAC2016\cite{MOPAC2016}. For better comparison enthalpy contributions and zero point energies where subtracted from the computed heat of formation.
Predictions with Paulings model Eq.~\eqref{eq:pauling} use electronegativities as reported by Pauling\cite{pauling}.

\section*{Supplementary Information Available}
Optimized parameters and predicted binding energies for double bonds. Additional data used for the derivation of our model. Total energies of single and double bonded molecules presented in this study.

\section*{Data}
Example input files and pseudopotentials to calculate alchemical atomic energies; binding energies, total energies and optimized structures in xyz-format calculated at PBE0/def2-tzvp level and code used to optimize the parameters in the different versions of our model are available at \url{https://doi.org/10.5281/zenodo.7421901}.

\section*{Acknowledgements}
We acknowledge discussions with M. Meuwly, M. Bragato and P. Marquetand, as well
as support from the European Research Council (ERCCoG Grant QML). This project has received funding
from the European Union's Horizon 2020 research and
innovation programme under Grant Agreement \#772834.
%Part of this research was performed while GvR was visiting the Institute for Pure and Applied Mathematics (IPAM), which is supported by the National Science Foundation (Grant No. DMS-1925919).

\bibliography{ref}

\end{document}

\title{Supplementary Information: From quantum alchemy to Hammett's equation:
Covalent bonding from  atomic energy partitioning}

\author{Michael J. Sahre}
\affiliation{University of Vienna, Faculty of Physics, Kolingasse 14-16, 1090 Vienna, Austria}
\affiliation{University of Vienna, Vienna Doctoral School in Chemistry (DoSChem), W\"ahringer Str. 42, 1090 Vienna, Austria.}
\author{Guido Falk von Rudorff}
\affiliation{University Kassel,
Department of Chemistry, 
Heinrich-Plett-Str.40,34132 Kassel, Germany}
\author{O. Anatole von Lilienfeld}
\email{anatole.vonlilienfeld@utoronto.ca}
\affiliation{Vector Institute for Artificial Intelligence, Toronto, ON, M5S 1M1, Canada}
\affiliation{Departments of Chemistry, Materials Science and Engineering, and Physics, University of Toronto, St. George Campus, Toronto, ON, Canada}
\affiliation{Machine Learning Group, Technische Universit\"at Berlin and Institute for the Foundations of Learning and Data, 10587 Berlin, Germany}

\maketitle

\newpage

\section*{Bond order}
\FloatBarrier

\begin{figure*}[ht]
\centering
\includegraphics[width=1.0\textwidth]{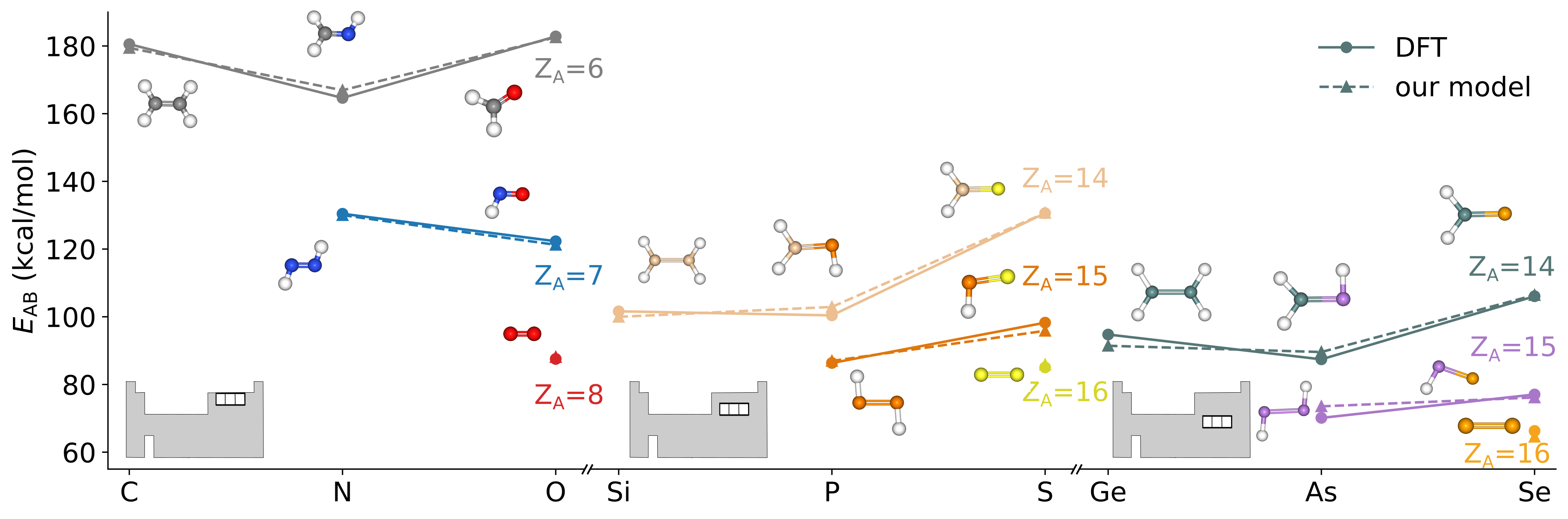}
\caption{Calculated bond dissociation energies of double bonds from density functional theory (PBE0/def2-TZVP) and our quantum alchemy based chemical bond model (Eq.~(2)).
MAEs with respect to DFT amount to 0.9, 1.3 and 2.0\,kcal/mol for the second, third and fourth row, respectively.}
\label{fig:bde_double}
\end{figure*}

\begin{table}
\centering
\caption{Coefficients and MAEs of our model Eq. (2) for  different rows $n$ after refitting to $\Delta E^\text{TS}$. $b$ is scaled such that $b Z_\text{A} Z_\text{B}$ is given in kcal/mol if $Z_\text{A}$ and $Z_\text{B}$ are given in atomic units.}
\begin{tabular}{c c c c c}
 $n$ & $a$ (kcal/mol) & $b$ (630/$a_0$) & $c$ (kcal/mol) & MAE (kcal/mol) \\
  \hline
2 & \ 439.13 & 24.3984 & 4.7284 & 0.9 \\
3 & \ 740.44 & 18.8836 & 3.2396 & 1.3 \\
4 & 2475.12 & 14.2104 & 1.8714 & 2.0 \\
\end{tabular}
\label{tab:coeff_TS}
\end{table}

\FloatBarrier
\newpage
\section*{Electronic configuration}

\begin{figure*}[ht]
\centering
\includegraphics[width=1.0\textwidth]{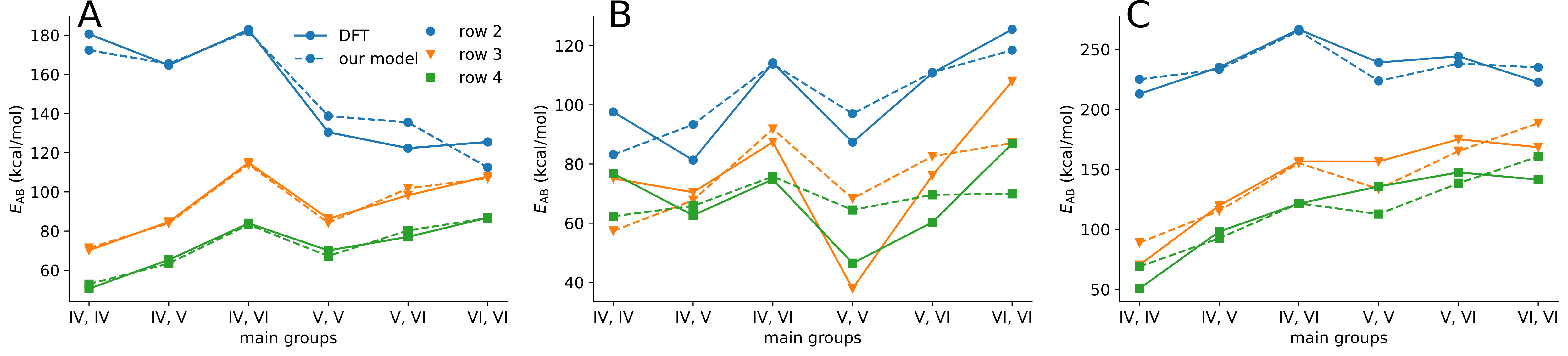}
\caption{Energy differences between double bonded molecules A=B and their fragments A, B with A and B from main groups IV, V, VI in the same row of the periodic table for various electronic states. Predictions are obtained from DFT (PBE0/def2-tzvp) and our model (Eq.~(2)). All molecules and fragments are either in their electronic ground state (Panel A), the energetically lowest triplet state (Panel B) or the lowest singlet state (Panel C).}
\label{fig:bde_double_GS_TT_SS}
\end{figure*}

\FloatBarrier

\begin{table}
\centering
\caption{Coefficients and MAEs of our model Eq. (2) for  different rows $n$ after refitting to $\Delta E^\text{GS}$. $b$ is scaled such that $b Z_\text{A} Z_\text{B}$ is given in kcal/mol if $Z_\text{A}$ and $Z_\text{B}$ are given in atomic units.}
\begin{tabular}{c c c c c}
 $n$ & $a$ (kcal/mol) & $b$ (630/$a_0$) & $c$ (kcal/mol) & MAE (kcal/mol) \\
  \hline
2 & 367.73 & 19.7386 & 3.9376 & 7.4 \\
3 & 371.53 & 12.4697 & 2.2692 & 1.6 \\
4 & 874.76 & \ 6.7190 & 0.9318 & 1.9 \\
\end{tabular}
\label{tab:coeff_GS}
\end{table}

\begin{table}
\centering
\caption{Coefficients and MAEs of our model Eq. (2) for  different rows $n$ after refitting to $\Delta E^\text{TT}$. $b$ is scaled such that $b Z_\text{A} Z_\text{B}$ is given in kcal/mol if $Z_\text{A}$ and $Z_\text{B}$ are given in atomic units.}
\begin{tabular}{c c c c c}
 $n$ & $a$ (kcal/mol) & $b$ (630/$a_0$) & $c$ (kcal/mol) & MAE (kcal/mol) \\
  \hline
2 & \ 89.83 & 6.4716 & 1.7297 & \ 7.3 \\
3 & 288.38 & 9.7959 & 1.7877 & 13.8 \\
4 & 758.64 & 4.8042 & 0.6495 & 10.5 \\
\end{tabular}
\label{tab:coeff_TT}
\end{table}

\begin{table}
\centering
\caption{Coefficients and MAEs of our model Eq. (2) for  different rows $n$ after refitting to $\Delta E^\text{SS}$. $b$ is scaled such that $b Z_\text{A} Z_\text{B}$ is given in kcal/mol if $Z_\text{A}$ and $Z_\text{B}$ are given in atomic units.}
\begin{tabular}{c c c c c}
 $n$ & $a$ (kcal/mol) & $b$ (630/$a_0$) & $c$ (kcal/mol) & MAE (kcal/mol) \\
  \hline
2 & 333.59 & 17.6638 & 4.0305 & \ 8.2 \\
3 & \ 80.28 & \ 8.2754 & 1.7258 & 12.9 \\
4 & \ \ 0.00 & \ 3.4398 & 0.5523 & 12.6 \\
\end{tabular}
\label{tab:coeff_SS}
\end{table}

\FloatBarrier

\begin{figure*}[ht]
\centering
\includegraphics[width=1.0\textwidth]{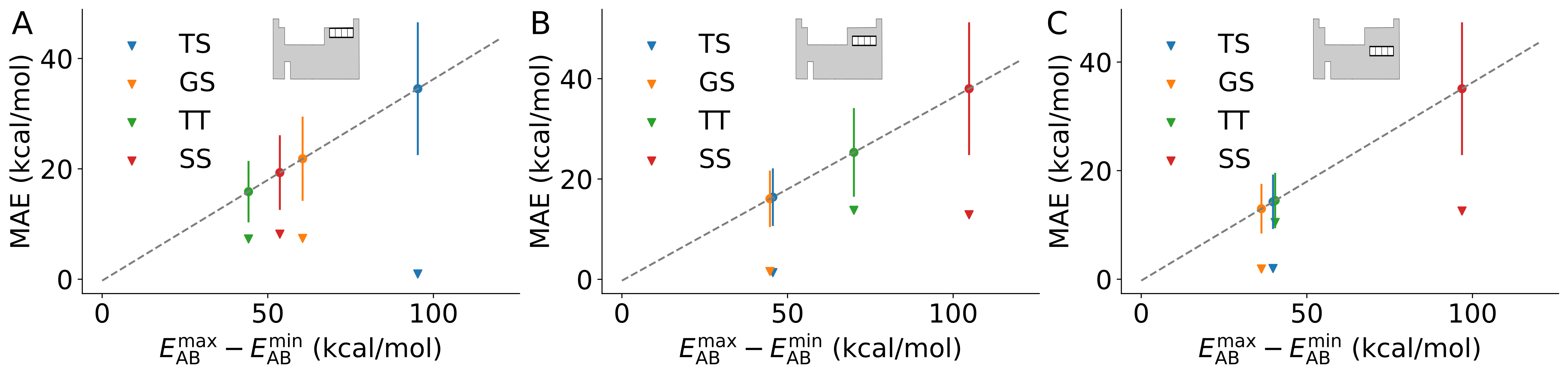}
\caption{The performance of our model after re-calibration to binding energy differences of various electronic states in comparison to a fit to random data points from a uniform distribution.
Triangles represent MAEs as a function of binding energy range after fitting our model to energy differences between fragments and double bonded molecules as triplets and singlets (TS), in their groundstate (GS), in the lowest triplet states (TT) and the lowest singlet states (SS). The circles with error bars show MAEs and standard deviation after fitting to random data points in the same energy ranges. Panels A, B, C show results for combinations of elements from main groups IV, V and VI for the second, third and fourth row of the periodic table, respectively.}
\label{fig:bde_double_error_all}
\end{figure*}

\FloatBarrier

\section*{Methods - Construction of the model}
\begin{figure}[H]
\centering
\includegraphics[width=0.7\textwidth]{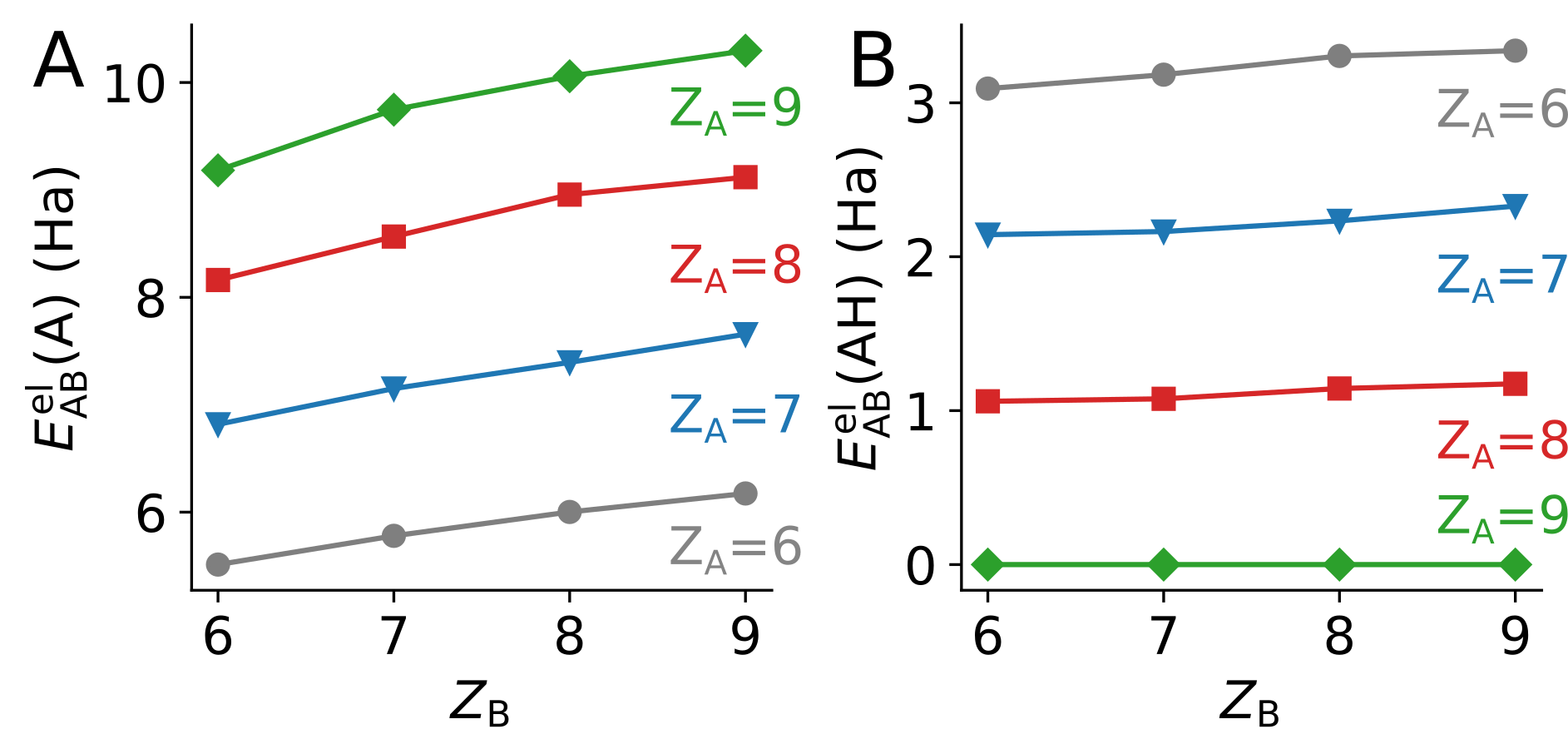}
\caption{Atomic electronic binding energies (see Eq.~(10)) of atom A as a function of the nuclear charge $Z_\text{B}$ of the binding partner for carbon, nitrogen, oxygen and fluorine (Panel A) and the cumulated atomic binding energies of the hydrogens attached to the different atoms A (Panel B).}
\label{fig:alchemical_energies}
\end{figure}

\begin{figure}[H]
\centering
\includegraphics[width=0.7\textwidth]{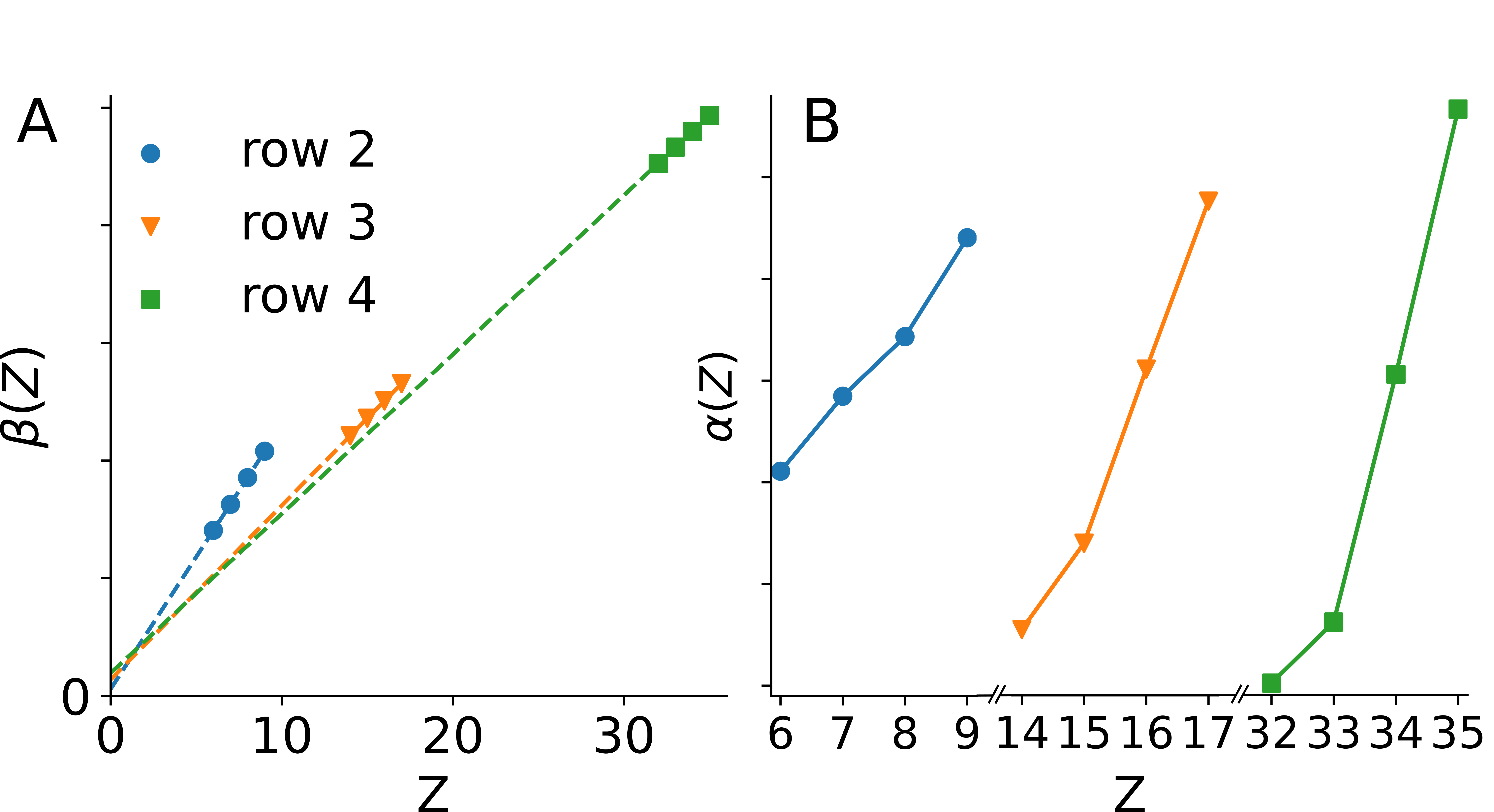}
\caption{Optimized parameters $\beta$ (Panel A) and $\alpha$ (Panel B) of Eq.~(12) as a function of $Z$ after a fit to binding energies calculated with DFT (PBE0/def2-TZVP).}
\label{fig:alpha_beta}
\end{figure}

\begin{figure}[H]
\centering
\includegraphics[width=0.7\textwidth]{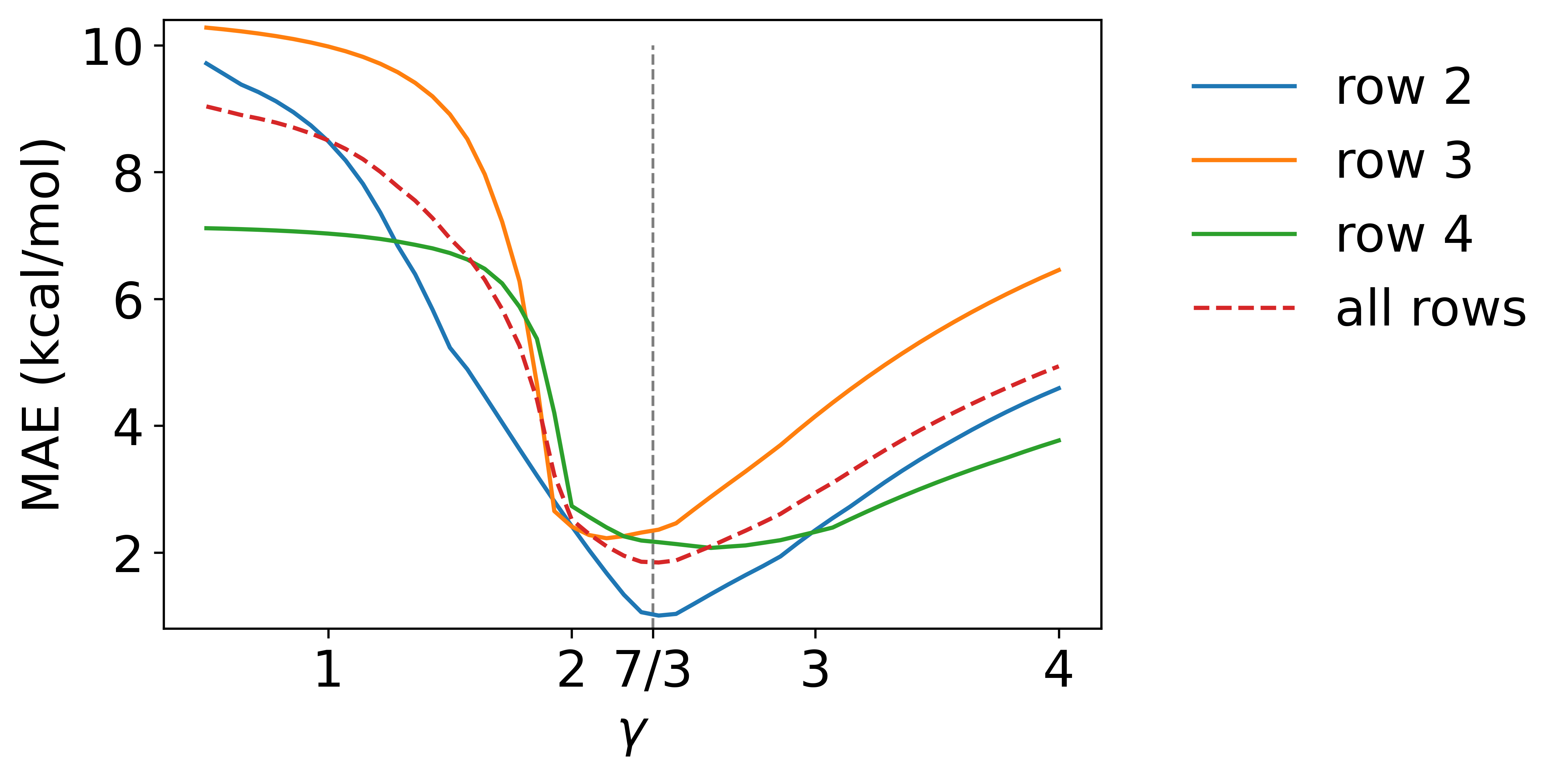}
\caption{The mean absolute error for different values of the exponent $\gamma$ in Eq.~(18). The curves labeled row 2, row 3 and row 4 show the MAE if the parameters in Eq.~(18) are optimized for each row individually. The dotted red curve shows the MAE for all three rows combined if $\gamma$ is kept the same for all rows.}
\label{fig:gamma}
\end{figure}

\begin{table}[H]
\centering
\caption{The average bond length $\bar{d}$ of A-B (in Bohr) and the optimized parameters $(\alpha, \beta)$ of Eq.~(12) for the elements from main group IV to VII for different rows $n$ of the periodic table. The unit of $\alpha$ is kcal/mol and $\beta$ is given as $630 \cdot e a_0^{-1}$ such that the unit of $\beta_\text{A} Z_\text{A}$ is also kcal/mol if $Z_\text{A}$ is provided in atomic units.}
\begin{tabular}{c c c c c c}
$n$ & $\bar{d} (a_0)$ & IV & V & VI & VII\\ 
\hline
2 & 2.69 & (55.39, 702.85) & (92.26, 813.36) & (121.54, 926.70) & (170.16, 1039.03)\\
3 & 4.01 & (-22.70, 1102.97) & (19.71, 1177.84) & (105.29, 1251.12) & (187.89, 1324.77)\\
4 & 4.46 & (-49.11, 2261.20) & (-19.08, 2330.51) & (102.73, 2397.54) & (233.19, 2464.38)\\
\label{tab:coeff_SRL}
\end{tabular}
\end{table}

\section*{Trends and dependence on period - Eq.(7)}
The $Z^{7/3}$ terms in our model can be approximated by a second order Taylor expansion as
\begin{equation}\label{eq:taylor_Z}
	Z^{7/3} \approx Z_0^{7/3} + \frac{7}{3} Z_0^{4/3} (Z-Z_0) + \frac{14}{9} Z_0^{1/3} (Z-Z_0)^2
\end{equation}
with $Z_0$ being a suitable reference nuclear charge.
Insertion of this expression into our model leads to
\begin{equation}\label{eq:taylor_Eb1}
	\begin{split}
		E_\text{AB} &= a - b Z_\text{A} Z_\text{B} + c\left( \frac{4}{9} Z_0^{7/3} - \frac{7}{9} Z_0^{4/3} (Z_\text{A} + Z_\text{B}) + \frac{14}{9} Z_0^{1/3} \left( Z_\text{A}^2 + Z_\text{B}^2 \right) \right) \\
		&= \underbrace{a + \frac{4}{9} Z_0^{7/3} c}_{=\xi} - \underbrace{\frac{7}{9} Z_0^{4/3} c}_{= \eta} (Z_\text{A} + Z_\text{B}) - b Z_\text{A} Z_\text{B} + \underbrace{\frac{14}{9} Z_0^{1/3} c}_{=\mu} \left( Z_\text{A}^2 + Z_\text{B}^2 \right).
	\end{split}
\end{equation}

The quadratic terms can be written as
\begin{equation}
    \begin{split}
         Z_\text{A}^2 + Z_\text{B}^2 &= Z_\text{A}^2 + Z_\text{B}^2 + 2 Z_\text{A} Z_\text{B} - Z_\text{A} Z_\text{B}\\
         &= (Z_\text{A} - Z_\text{B})^2 + 2 Z_\text{A} Z_\text{B}
    \end{split}
\end{equation}
leading to 
\begin{equation}\label{eq:taylor_Eb2}
	\begin{split}
		E_\text{AB}
		&= \xi - \eta (Z_\text{A} + Z_\text{B})
		- b Z_\text{A} Z_\text{B} + \mu \left( [Z_\text{A} - Z_\text{B}]^2 + 2 Z_\text{A} Z_\text{B} \right) \\
		&= \xi - \eta (Z_\text{A} + Z_\text{B}) + (2 \mu - b) Z_\text{A} Z_\text{B} + \mu \left( Z_\text{A} - Z_\text{B}\right)^2.
	\end{split}
\end{equation}
The energy differences $\Delta E = E_\text{AB} - E_\text{AC}$ between A-B and A-C is then
\begin{equation}
\begin{split}
    \Delta E &= -\eta \left(Z_\text{B} - Z_\text{C}\right) + \kappa Z_\text{A} \left(Z_\text{B} - Z_\text{C}\right) 
    + \mu \left( \left( Z_\text{A} - Z_\text{B}\right)^2 - \left( Z_\text{A} - Z_\text{C}\right)^2 \right)
\end{split}
\end{equation}
with $\kappa = (2 \mu - b)$. 

% Values for $\eta, \kappa, \mu$ are shown in TABLE~\ref{tab:coeffs_ediff_intial} for $Z_0$ being the mean nuclear charge of the considered elements within one row.
% \begin{table}
% \centering
% \caption{Coefficients and MAEs for model Eq. \eqref{eq:SRL*} for row $n = 2, 3, 4$.}
% \begin{tabular}{ c c c c c}
% n & $Z_0$ & $\eta$ & $\kappa$ & $\mu$ \\ 
% \hline
% 2 & 7.5 & 22.7 & 1.6 & 6.0 \\
% 3 & 15.5 & 45.0 & 2.8 & 5.8 \\
% 4 & 33.5 & 68.3 & 2.0 & 4.1 \\
% \end{tabular}
% \label{tab:coeffs_ediff_intial}
% \end{table}

To account for the dependence of $\eta, \kappa$ and $\mu$ on the row or principal quantum number $n$ we make the ansatz $\eta = \tilde{\eta} (n - 1)$, $\kappa = (n-3)^2 - \tilde{\kappa} $, while we choose $\mu$ to be independent from the prinicipal quantum number.
Non-linear least squares optimization of $\tilde{\eta}$, $\tilde{\kappa}$ and $\mu$ with respect to $\Delta E$ in the resulting expression
\begin{equation}
\begin{split}
    \Delta E &= -\tilde{\eta} (n - 1) \left(Z_\text{B} - Z_\text{C}\right) + [(n-3)^2 - \tilde{\kappa}] Z_\text{A} \left(Z_\text{B} - Z_\text{C}\right) + \mu \left( \left( Z_\text{A} - Z_\text{B}\right)^2 - \left( Z_\text{A} - Z_\text{C}\right)^2 \right).
\end{split}
\end{equation}
gives $\tilde{\eta} = 28$, $\tilde{\kappa} = 8.5$ and $\mu = 6$, so that
\begin{equation}
\begin{split}
    \Delta E &= -28 (n - 1) \left(Z_\text{B} - Z_\text{C}\right) - [8.5 + (n-3)^2] Z_\text{A} \left(Z_\text{B} - Z_\text{C}\right) + 6 \left( \left( Z_\text{A} - Z_\text{B}\right)^2 - \left( Z_\text{A} - Z_\text{C}\right)^2 \right)
\end{split}
\end{equation}
% \begin{equation}
% \begin{split}
%     \Delta E_\text{ABC} &= -28 (n-1) \Delta Z_\text{BC} - ([n-3]^2 - 3.5) Z_\text{A} \Delta Z_\text{BC} \\
%     &+ 6 \left( \Delta Z_\text{AB}^2 - \Delta Z_\text{AC}^2 \right).
% \end{split}
% \end{equation}
or alternatively
\begin{equation}
\begin{split}
    \Delta E &=  -28 (n-1) \Delta Z - (8.5 + (n-3)^2) Z_\text{A} \Delta Z + 6 (Z_\text{B}^2 - Z_\text{C}^2)
\end{split}
\end{equation}
with $\Delta Z = Z_\text{B} - Z_\text{C}$.

\newpage

\FloatBarrier

\section*{Total Energies}
\begin{table}
\caption{Total energies (in Hartree) of single bonded molecules and respective fragments after homolytic bond cleavage optimized with PBE0/def2-tzvp.}
\begin{tabular}{llllll}
\toprule
compound & energy & compound & energy & compound & energy \\
\midrule
C$_2$H$_6$ & -79.750204 & Cl$_2$ & -920.089409 & BeHCH$_3$ & -55.172389 \\
CH$_3$NH$_2$ & -95.777291 & Ge$_2$H$_6$ & -4157.072524 & BeHNH$_2$ & -71.250843 \\
CH$_3$OH & -115.637262 & GeH$_3$AsH$_2$ & -4315.341268 & BeHOH & -91.138621 \\
CH$_3$F & -139.653810 & GeH$_3$SeH & -4480.425039 & BeHF & -115.173176 \\
N$_2$H$_4$ & -111.784922 & GeH$_3$Br & -4652.430337 & Li & -7.467048 \\
NH$_2$OH & -131.622414 & As$_2$H$_4$ & -4473.605476 & BeH & -15.223816 \\
NH$_2$F & -155.625694 & AsH$_2$SeH & -4638.681646 & BH$_2$ & -25.893672 \\
H$_2$O$_2$ & -151.453106 & AsH$_2$Br & -4810.679452 & CH$_3$ & -39.796931 \\
OHF & -175.439419 & H$_2$Se$_2$ & -4803.749098 & NH$_2$ & -55.832413 \\
F$_2$ & -199.408535 & SeHBr & -4975.731476 & OH & -75.683320 \\
SiH$_3$SiH$_3$ & -582.338252 & Br$_2$ & -5147.706716 & F & -99.673999 \\
SiH$_3$PH$_2$ & -633.603800 & BH$_2$CH$_3$ & -65.866689 & SiH$_3$ & -291.107812 \\
SiH$_3$SH & -689.861429 & BH$_2$NH$_2$ & -81.963033 & PH$_2$ & -342.378482 \\
SiH$_3$Cl & -751.278691 & BH$_2$OH & -101.826384 & SH & -398.609973 \\
P$_2$H$_4$ & -684.856580 & BH$_2$F & -125.840622 & Cl & -459.995117 \\
PH$_2$SH & -741.101414 & LiCH$_3$ & -47.336002 & GeH$_3$ & -2078.478952 \\
PH$_2$Cl & -802.505076 & LiNH$_2$ & -63.410044 & AsH$_2$ & -2236.760002 \\
H$_2$S$_2$ & -797.331489 & LiOH & -83.307236 & SeH & -2401.827158 \\
SHCl & -858.715428 & LiF & -107.347917 & Br & -2573.810130 \\
\bottomrule
\end{tabular}
\end{table}

\begin{table}
\caption{Total energies (in Hartree) of double bonded molecules and respective fragments after homolytic bond cleavage optimized with PBE0/def2-tzvp.}
\begin{tabular}{llllll}
\toprule
compound & energy singlet & energy triplet & compound & energy singlet & energy triplet\\
\midrule
C$_2$H$_4$ & -78.512205 & -78.380516 & GeH$_2$Se & -4479.211917 & -4479.162320 \\
CH$_2$NH & -94.550022 & -94.417719 & H$_2$As$_2$ & -4472.399893 & -4472.362319 \\
CH$_2$O & -114.418666 & -114.309740 & AsHSe & -4637.469705 & -4637.443125 \\
H$_2$N$_2$ & -110.558763 & -110.490383 & Se$_2$ & -4802.511423 & -4802.544133 \\
NHO & -130.385650 & -130.367261 & CH$_2$ & -39.087164 & -39.112821 \\
O$_2$ & -150.170231 & -150.230448 & NH & -55.089690 & -55.175865 \\
Si$_2$H$_4$ & -581.083609 & -581.041531 & O & -74.908473 & -75.015640 \\
SiH$_2$PH & -632.371719 & -632.324131 & SiH$_2$ & -290.486074 & -290.461159 \\
SiH$_2$S & -688.639863 & -688.571246 & PH & -341.695472 & -341.751186 \\
H$_2$P$_2$ & -683.639421 & -683.562517 & S & -397.905241 & -397.971393 \\
PHS & -739.878544 & -739.843432 & GeH$_2$ & -2077.875509 & -2077.840458 \\
S$_2$ & -796.077755 & -796.114103 & AsH & -2236.092207 & -2236.144296 \\
Ge$_2$H$_4$ & -4155.831310 & -4155.802722 & Se & -2401.143488 & -2401.203114 \\
GeH$_2$AsH & -4314.123511 & -4314.084186 &  &  &  \\
\bottomrule
\end{tabular}
\end{table}